\documentclass[preprint]{aastex62}
\pdfoutput=1 %for arXiv submission

\usepackage{amsmath,amstext}
\usepackage{multirow}

\newcommand{\gray}{gamma-ray }
\newcommand{\agile}{{\it AGILE }}

\shorttitle{\agile gamma-ray sources coincident with cosmic neutrino events}
\shortauthors{Lucarelli et al.}

\turnoffediting

\begin{document}

\title{\agile detection of gamma-ray sources coincident with cosmic neutrino events}

%%%%%%%%%%%%%%%%%%%%%%%%%%%%%%%%%%%%%%%%%%%%%%%%%%%%%%%%%%%%%%%%%%%%%%%%%%%%%
%Authors
\correspondingauthor{Fabrizio Lucarelli, Marco Tavani}
\email{fabrizio.lucarelli@ssdc.asi.it, marco.tavani@iaps.inaf.it}

%\author[0000-0002-6311-764X]{F. Lucarelli}
\author{F. Lucarelli}
\affiliation{ASI Space Science Data Center (SSDC), Via del Politecnico snc, I-00133 Roma, Italy}
\affiliation{INAF--OAR, via Frascati 33, I-00078 Monte Porzio Catone (Roma), Italy}

%\author[0000-0003-2893-1459]{M. Tavani}
\author{M. Tavani}
\affiliation{INAF/IAPS--Roma, Via del Fosso del Cavaliere 100, I-00133 Roma, Italy}
\affiliation{Univ. ``Tor Vergata", Via della Ricerca Scientifica 1, I-00133 Roma, Italy}

%\author[0000-0002-9332-5319]{G. Piano}
\author{G. Piano}
\affiliation{INAF/IAPS--Roma, Via del Fosso del Cavaliere 100, I-00133 Roma, Italy}

%\author[0000-0001-6347-0649]{A. Bulgarelli}
\author{A. Bulgarelli}
\affiliation{INAF/IASF–Bologna, Via Gobetti 101, I-40129 Bologna, Italy}

%\author[0000-0002-4700-4549]{I. Donnarumma}
\author{I. Donnarumma}
\affiliation{Agenzia Spaziale Italiana (ASI), Via del Politecnico snc, I-00133 Roma, Italy}

%\author[0000-0003-3455-5082]{F. Verrecchia}
\author{F. Verrecchia}
\affiliation{ASI Space Science Data Center (SSDC), Via del Politecnico snc, I-00133 Roma, Italy}
\affiliation{INAF--OAR, via Frascati 33, I-00078 Monte Porzio Catone (Roma), Italy}

%\author[0000-0001-6661-9779]{C. Pittori}
\author{C. Pittori}
\affiliation{ASI Space Science Data Center (SSDC), Via del Politecnico snc, I-00133 Roma, Italy}
\affiliation{INAF--OAR, via Frascati 33, I-00078 Monte Porzio Catone (Roma), Italy}

%Team
\author{L. A. Antonelli}
\affiliation{INAF--OAR, via Frascati 33, I-00078 Monte Porzio Catone (Roma), Italy}

\author{A. Argan}
\affiliation{INAF/IAPS--Roma, Via del Fosso del Cavaliere 100, I-00133 Roma, Italy}

\author{G. Barbiellini}
\affiliation{Dipartimento di Fisica, Università di Trieste and INFN, via Valerio 2,I-34127 Trieste, Italy}

\author{P. Caraveo}
\affiliation{INAF/IASF--Milano, via E.Bassini 15, I-20133 Milano, Italy}

\author{M. Cardillo}
\affiliation{INAF/IAPS--Roma, Via del Fosso del Cavaliere 100, I-00133 Roma, Italy}

\author{P. W. Cattaneo}
\affiliation{INFN--Pavia, Via Bassi 6, I-27100 Pavia, Italy}

\author{A. Chen}
\affiliation{University of Witwatersrand, Johannesburg, South Africa}

\author{S. Colafrancesco}
\affiliation{University of Witwatersrand, Johannesburg, South Africa}
\affiliation{INAF--OAR, via Frascati 33, I-00078 Monte Porzio Catone (Roma), Italy}

\author{E. Costa}
\affiliation{INAF/IAPS--Roma, Via del Fosso del Cavaliere 100, I-00133 Roma, Italy}
\affiliation{Agenzia Spaziale Italiana (ASI), Via del Politecnico snc, I-00133 Roma, Italy}

\author{E. Del Monte}
\affiliation{INAF/IAPS--Roma, Via del Fosso del Cavaliere 100, I-00133 Roma, Italy}

\author{G. Di Cocco}
\affiliation{INAF/IASF--Bologna, Via Gobetti 101, I-40129 Bologna, Italy}

\author{A. Ferrari}
\affiliation{CIFS, c/o Physics Department, University of Turin, via P. Giuria 1, I-10125 Torino, Italy}

\author{V. Fioretti}
\affiliation{INAF/IASF–Bologna, Via Gobetti 101, I-40129 Bologna, Italy}

\author{M. Galli}
\affiliation{INAF/IASF–Bologna, Via Gobetti 101, I-40129 Bologna, Italy}

\author{P. Giommi}
\affiliation{Agenzia Spaziale Italiana (ASI), Via del Politecnico snc, I-00133 Roma, Italy}

\author{A. Giuliani}
\affiliation{INAF/IASF--Milano, via E.Bassini 15, I-20133 Milano, Italy}

\author{P. Lipari}
\affiliation{INFN--Roma Sapienza, Piazzale Aldo Moro 2, 00185 Roma, Italy}

\author{F. Longo}
\affiliation{Dip. di Fisica, Universita’ di Trieste and INFN, Via Valerio 2, I-34127 Trieste, Italy}

\author{S. Mereghetti}
\affiliation{INAF/IASF--Milano, via E.Bassini 15, I-20133 Milano, Italy}

\author{A. Morselli}
\affiliation{INFN--Roma Tor Vergata, via della Ricerca Scientifica 1, 00133 Roma, Italy}

\author{F. Paoletti}
\affiliation{East Windsor RSD, 25a Leshin Lane, Hightstown, NJ 08520, USA}
\affiliation{INAF/IAPS--Roma, Via del Fosso del Cavaliere 100, I-00133 Roma, Italy}

\author{N. Parmiggiani}
\affiliation{INAF/IASF--Bologna, Via Gobetti 101, I-40129 Bologna, Italy}

\author{A. Pellizzoni}
\affiliation{INAF -- Osservatorio Astronomico di Cagliari, via della Scienza 5, I-09047 Selargius (CA), Italy}

\author{P. Picozza}
\affiliation{INFN--Roma Tor Vergata, via della Ricerca Scientifica 1, I-00133 Roma, Italy}

\author{M. Pilia}
\affiliation{INAF -- Osservatorio Astronomico di Cagliari, via della Scienza 5, I-09047 Selargius (CA), Italy}

\author{A. Rappoldi}
\affiliation{INFN--Pavia, Via Bassi 6, I-27100 Pavia, Italy}

\author{A. Trois}
\affiliation{INAF -- Osservatorio Astronomico di Cagliari, via della Scienza 5, I-09047 Selargius (CA), Italy}

\author{A. Ursi}
\affiliation{INAF/IAPS--Roma, Via del Fosso del Cavaliere 100, I-00133 Roma, Italy}

\author{S. Vercellone}
\affiliation{INAF -- Oss. Astron. di Brera, Via E. Bianchi 46, I-23807 Merate (LC), Italy}

\author{V. Vittorini}
\affiliation{INAF/IAPS--Roma, Via del Fosso del Cavaliere 100, I-00133 Roma, Italy}

%\author{....}
\collaboration{(The \agile Team)}
\noaffiliation

%%%%%%%%%%%%%%%%%%%%%%%%%%%%%%%%%%%%%%%%%%%%%%%%%%%%%%%%%%%%%%%%%%%%%%%%%%%%%

\begin{abstract}
The origin of cosmic neutrinos is still largely unknown. Using data 
obtained by the gamma-ray imager on board of the \agile satellite, we 
systematically searched for transient gamma-ray sources above 100 MeV that 
are temporally and spatially coincident with ten recent high-energy neutrino 
IceCube events. We find three \agile candidate sources that can be considered 
possible counterparts to neutrino events. Detecting 3 gamma-ray/neutrino 
associations out of 10 IceCube events is shown to be unlikely due to a chance 
coincidence. One of the sources is related to the BL Lac source TXS 0506+056. For 
the other two \agile gamma-ray sources there are no obvious known counterparts, and 
both Galactic and extragalactic origin should be considered.
\end{abstract}


%%%%%%%%%%%%%%%%%%%%%%%%%%%%%%%%%%%%%%%%%%%%%%%%%%%%%%%%%%%%%%%%%%%%
\section{Introduction}\label{intro}
%%%%%%%%%%%%%%%%%%%%%%%%%%%%%%%%%%%%%%%%%%%%%%%%%%%%%%%%%%%%%%%%%%%%
The discovery of a diffuse flux of cosmic neutrinos by the IceCube 
experiment~\citep{2013Sci...342E...1I, 2015PhRvL.115h1102A} %(see also~\cite{2018ApJ...853L...7A}) 
opened a new field of investigation in the context of neutrino astronomy 
(after the detections of the Sun and SN1987a). 
Energetic neutrinos of energies above 10 TeV can be produced in astrophysical 
beam dumps, where cosmic rays are accelerated in regions near compact 
objects or in shock fronts, and interact via proton-proton 
($p-p$) or proton-photon ($p-\gamma$) collisions with matter or radiation fields 
surrounding the central engine or within an ejected plasma flow (see~\cite{2017NatPh..13..232H} 
for a review). High-energy gamma-ray emission above the GeV is expected to be associated 
with these hadronic processes, with intensities that vary depending on source 
characteristics and environment~\citep{2017ARNPS..67...45M}.

No significant clustering of neutrinos above the expected background 
has been observed so far from any of the current experiments after 
several years of observations~\citep{2017ApJ...835..151A, 2017PhRvD..96h2001A}. 
Active Galactic Nuclei (AGNs) of the blazar category are considered as the main 
cosmic neutrino source candidates~\citep{1995APh.....3..295M}, although it has 
been suggested, based on the average properties, that they contribute only 
to a fraction of the observed diffuse flux~\citep{2017ApJ...835...45A}. A contribution 
from other types of active galaxies (starburst galaxies, 
radio-galaxies)~\citep{2006JCAP...05..003L, 2014PhRvD..89l3005B, 2018MNRAS.tmp..247T},
galaxy clusters/groups~\citep{2008ApJ...689L.105M, 2009ApJ...707..370K}, 
AGN winds~\citep{2016JCAP...12..012W, 2017A&A...607A..18L}, and Galactic sources (supernovae 
remnants expanding in dense molecular clouds, microquasars, hidden compact objects) 
should also be considered~\citep{2006APh....26..310V, 2005ApJ...631..466B, 2014ApJ...780...29S, 
Anchordoqui:2013dnh, 2016PhRvD..93a3009A}.

Observation of transient gamma-ray emission, spatially and temporally 
compatible with the IceCube neutrinos, is then crucial to identify their 
electro-magnetic (e.m.) counterparts. Since April 2016, the IceCube Collaboration 
is alerting the astronomical community almost in real time whenever a 
single-track high-energy starting event (HESE) or an extremely 
high-energy (EHE) \edit1{through-going track event, with an energy higher}
than several hundred TeV, is detected~\citep{2017APh....92...30A}. The implementation of 
the IceCube alert system with the possibility of fast follow-up 
observations by several space- and ground-based instruments allows 
a global search for this association. On September 2017 a first significant 
association (at the level of 3$\sigma$) was announced: the gamma-ray flaring 
blazar of the BL Lac class, TXS 0506+056, was identified as a likely e.m. 
counterpart to the IceCube event IC-170922~\citep{2018Sci...361.1378I}.
Furthermore, from the analysis of archival data, an excess of VHE 
neutrinos from the direction of the same source has been 
also detected in 2014/2015~\citep{2018Sci...361..147I}. TXS 0506+056 has thus 
suggested as the first \edit1{extragalactic} neutrino point-like source ever detected.

We report here on a systematic search for \agile transient gamma-ray 
counterparts to the IceCube HESE/EHE events announced through the GCN/AMON system. 
The paper is organized as follows: in Section~\ref{agile_icecube-alerts} 
we present the results of the systematic search for gamma-ray sources, in coincidence 
with neutrino events, automatically detected by the \agile {\it Quick Look} transient 
detection system. The level of {\it AGILE}/IceCube correlation for some significant 
gamma-ray detections found in the search, is then evaluated estimating 
the probability to be accidental using the \agile False Alarm Rate (FAR) 
computed through the method discussed in Appendix~\ref{FAR}. 
In Section~\ref{counterparts}, we further investigated 
the common {\it AGILE}/IceCube detections, and we explore the possible e.m. 
counterpart candidates using the cross-catalog search tools available 
from the ASI Space Science Data Center\footnote{http://www.ssdc.asi.it}. 
Finally, in Section~\ref{discussion}, we discuss the astrophysical implications 
of the \agile observations.

\vspace{0.5cm}

%%%%%%%%%%%%%%%%%%%%%%%%%%%%%%%%%%%%%%%%%%%%%%%%%%%%%%%%%%%%%%%%%%%%
\section{The \agile satellite search for gamma-ray counterparts to 
IceCube neutrino events.}
\label{agile_icecube-alerts}
%%%%%%%%%%%%%%%%%%%%%%%%%%%%%%%%%%%%%%%%%%%%%%%%%%%%%%%%%%%%%%%%%%%%
The \agile {(\it Astro–rivelatore Gamma a Immagini Leggero}) satellite 
monitors cosmic gamma-ray sources in the energy range from 30~MeV to 30~GeV~\citep{2009A&A...502..995T}. 
Since November 2009, the satellite scans the whole sky in spinning mode, being 
an all-sky detector for transient gamma-ray sources capable to expose about 80\% of 
the whole sky at any given time with good sensitivity and angular resolution to 
gamma-rays above 100~MeV.

In this observing mode, at the end of July 2016, the main instrument 
onboard of the satellite, the gamma-ray imager GRID, detected a gamma-ray transient 
(AGL J1418+0008) spatially and temporally consistent with the 
IceCube event IC-160731~\citep{2017ApJ...846..121L}. 
This detection was the result of the automatic and {\it Quick Look} (QL) search 
for gamma-ray transients above 100 MeV, daily performed over predefined 2-day 
integration time-bins of {\it AGILE}-–GRID data~\citep{2014ApJ...781...19B}.

Motivated by this first detection, we have explored the \agile QL database 
searching for other transient gamma-ray detections \replaced{spatially and temporally 
consistent with the reconstructed arrival directions of the 
IceCube HESE/EHE neutrino events observed since Apr. 2016.}{\edit1{with the 
following characteristics: 1) a centroid positionally compatible, 
within the \agile angular resolution, with the reconstructed 
arrival directions of the IceCube HESE/EHE neutrino events; and 
2) temporally occurring within a fixed search time window around 
the neutrino event time T$_0$.}} \added{Since Apr. 2016,} a total of 13 
neutrino events have been made public to 
date\footnote{https://gcn.gsfc.nasa.gov/amon\_hese\_events.html and 
https://gcn.gsfc.nasa.gov/amon\_ehe\_events.html} (see Appendix~\ref{hese_list} 
for the complete list); 10 events survive additional checks by the IceCube team. 
In this paper, we consider these 10 events as the basis for our study.

Typical IceCube HESE/EHE error circles are of order of $1^\circ$ in radius, 
and the mean \agile angular resolution measured from in-flight and 
calibration data is $\sim 1.5^\circ$ in the energy range 
100~MeV–-1~GeV~\citep{2015ApJ...809...60S}. Based on this, we have tried 
three values for the database search radius around the \added{\edit1{10 IceCube}} 
input sky positions ($1.0^\circ$, $1.5^\circ$, and $2.0^\circ$), which are 
within the range of the \agile point-spread function. 
\added{\edit1{Concerning the time window of interest, the astrophysics 
and timescales of the phenomena related to the emission of these 
extremely high-energy neutrinos and their likely correlated gamma-ray 
emission are still uncertain. Thus, based on the typical \agile sensitivity 
to a transient gamma-ray source, we consider the \agile QL detection 
{\it temporally} consistent with the neutrino event {\it if} occurs 
within a time interval of plus/minus 4~days around T$_0$.}} %$\Delta{T} = \pm 4$

\begin{table}[t!]
\begin{center}
\caption{The 3 \agile QL detections close in time and space to 
IceCube HESE/EHE neutrinos. Columns 2 to 5 show the main parameters 
of the corresponding IceCube event (event ID, neutrino event time T$_0$, 
best-fit reconstructed centroid position in Equatorial coordinates). 
Columns 6 to 9 show, respectively, the \agile gamma-ray flux (above 100~MeV) 
estimated over the QL 2-day integration time bin, the distance in time 
\replaced{$\delta{t}$}{\edit1{$\Delta{t}$}} from the QL detection centroid, 
the false alarm rate (FAR) expected for each detection\tablenotemark{a}, and 
the corresponding {\it post-trial} false alarm probability $P_i$.}
\label{table_agile_detections}
\vskip .3cm
\scriptsize
\begin{tabular}{l|c|c|c|c|c|c|c|c}
\hline\hline
\agile & IceCube & T$_0$ & R.A. (J2000) & Decl. (J2000) & $F_{\gamma}(E>100~MeV)$ &
\edit1{$\Delta{t}$} & FAR & $P_i$ \\
source & event & (MJD) & (deg) & (deg) & $\times 10^{-6} \,\rm (ph\, cm^{-2}\, s^{-1})$ & (days) 
& & {\it post-trial} \\
\hline
A & IC-160731 & 57600.079 & 214.544 & -0.3347 & $(1.8 \pm 0.7)$ & -2.0 & $5.9 \times 10^{-4}$ 
& $2.0 \times 10^{-3}$ \\
B & IC-170321 & 57833.314 & 98.3 & -15.02 & $(1.5 \pm 0.6)$ & -2.2 & $1.5 \times 10^{-3}$ 
& $5.7 \times 10^{-3}$ \\ 
C & IC-170922 & 58018.871 & 77.43 & 5.72 & $(1.7 \pm 0.7)$ & -2.8 & $1.0 \times 10^{-3}$ 
& $5.0 \times 10^{-3}$ \\
\hline
\end{tabular}
\end{center}
\tablenotetext{a}{See Appendix~\ref{FAR} for the details about the \agile FAR estimate.}
\end{table}

From the QL database mining, using the optimized $1.5^\circ$ search cone radius, we 
found 3 significant \agile detections which \replaced{are consistent in time and 
space}{\edit1{satisfy the temporal and spatial association 
criteria defined above in correspondence of}} the following 
three IceCube events: \replaced{IC-160731 (from now on \agile Source A), 
IC-170321 (Source B), and IC-170922 (Source C)}{\edit1{IC-160731, IC-170321, 
and IC-170922. From now on, we will indicate the corresponding three \agile 
detections as \agile Source A, Source B, and Source C, respectively.}} 
Table~\ref{table_agile_detections} shows the details about the 3 \agile QL 
detections, along with the main parameters of the closest IceCube 
detection (event ID, neutrino event time T$_0$, and best-fit reconstructed 
centroid position in Equatorial coordinates). In all cases, the gamma-ray position is 
within the \agile angular resolution ($1.5^\circ$) from the best-fit IceCube reconstructed arrival 
direction. The Table also shows the time difference \added{\edit1{$\Delta{t}$}} between 
T$_0$ and the center of the 2-day integration interval of the closest QL detection, and the 
corresponding gamma-ray flux above 100 MeV evaluated by means of the \agile maximum 
likelihood (ML) algorithm~\citep{2012A&A...540A..79B}. 

The first {\it AGILE}/IceCube event in Table~\ref{table_agile_detections} is 
AGL J1418+0008, whose detection was already reported in \cite{2017ApJ...846..121L}. 
The \agile Source B is reported here for the first time: this detection, 
with a gamma-ray flux above 100 MeV of 
$F=(1.5 \pm 0.6) \times 10^{-6} \, \rm ph \, cm^{-2} \, s^{-1}$, 
is temporally close (2 days prior) to the IceCube event 
which occurred on 21 Mar. 2017~\citep{2017GCN.20929....1B}. The last one, Source C, 
corresponds to the most recent IceCube-170922A event and is consistent with 
the gamma-ray activity from the blazar TXS 0506+056 as reported in 
\cite{2018Sci...361.1378I}.
\deleted{Figure 1 shows the distribution of all IceCube events in a 
Hammer-Aitoff projection of the sky in Galactic coordinates. 
All events appear well above the Galactic plane, except for one case 
(IC-170321) which shows a Galactic latitude of -10.75 degrees. 
The three neutrino events with an \agile possible transient 
counterpart, A, B, and C, are shown in red.} 
We notice that all the three events with an \agile nearby source detection 
belong to the extremely high-energy (EHE) event class with track-like 
characteristics~\citep{2017APh....92...30A} (see also Table~\ref{ICECUBE_ALERTS} 
in App.~\ref{hese_list}).

\begin{figure*}[!t]
\gridline{
          \leftfig{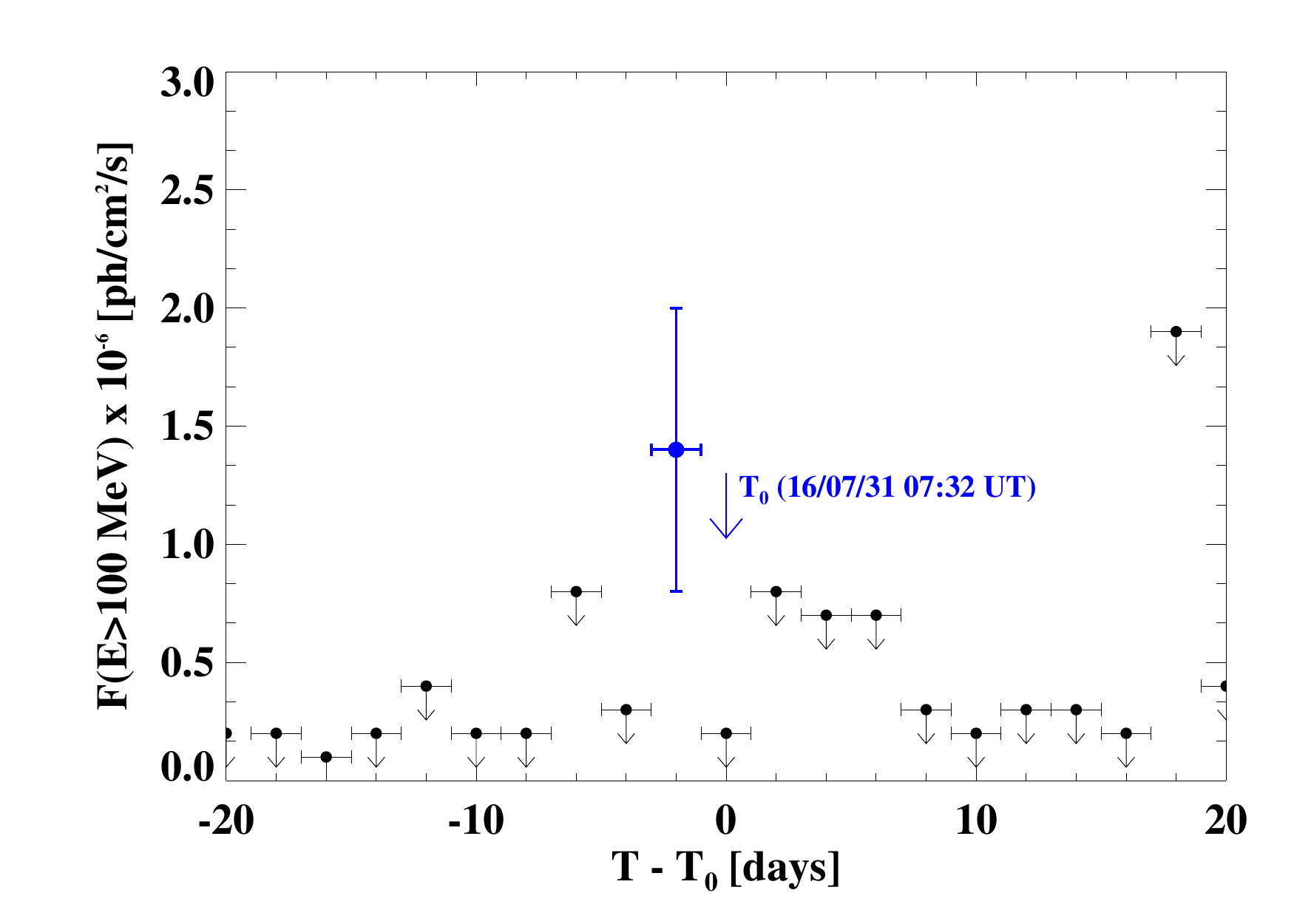}{0.5\textwidth}{}
          \rightfig{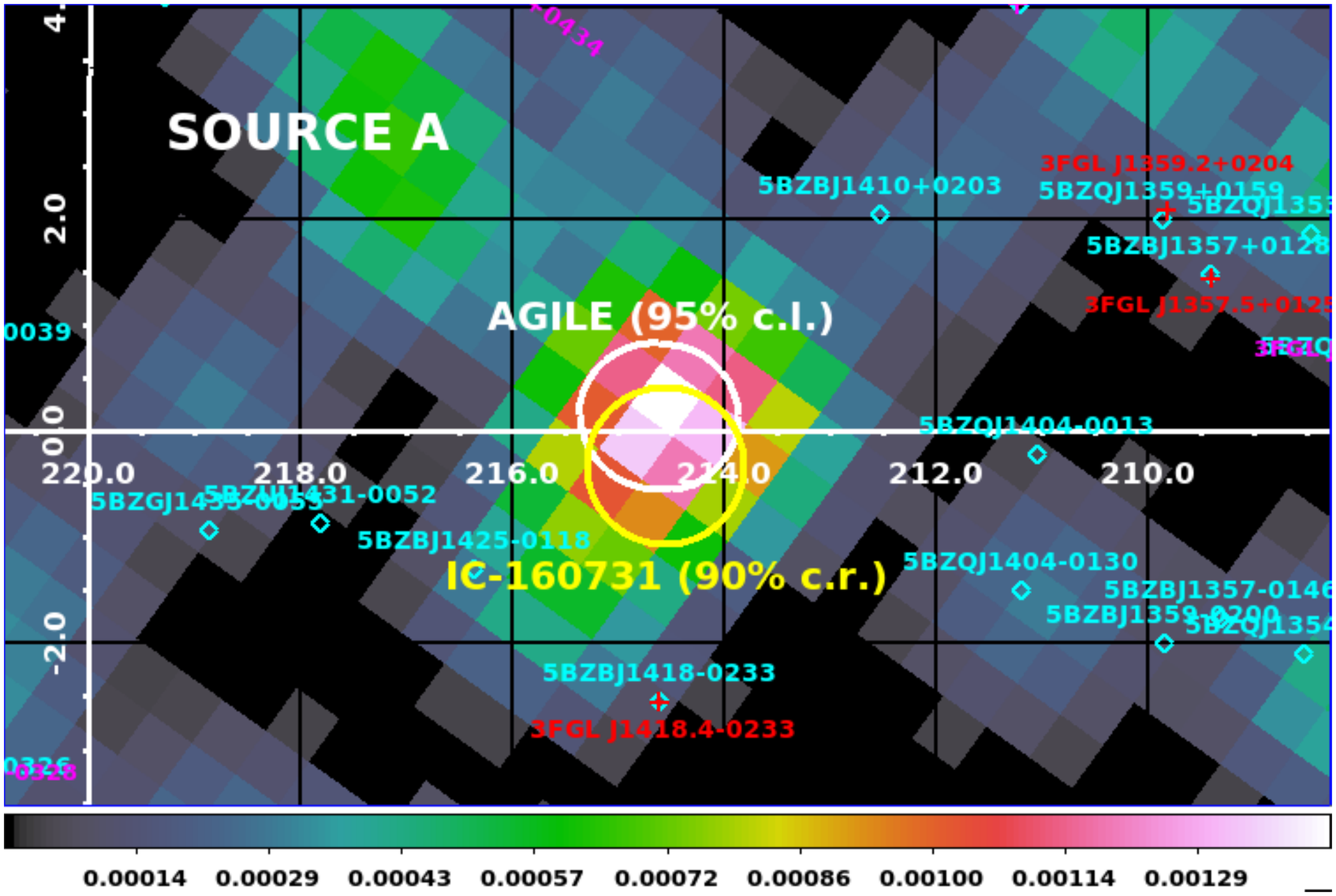}{0.5\textwidth}{}          
          }
\vspace{-1.2cm}
\gridline{
          \leftfig{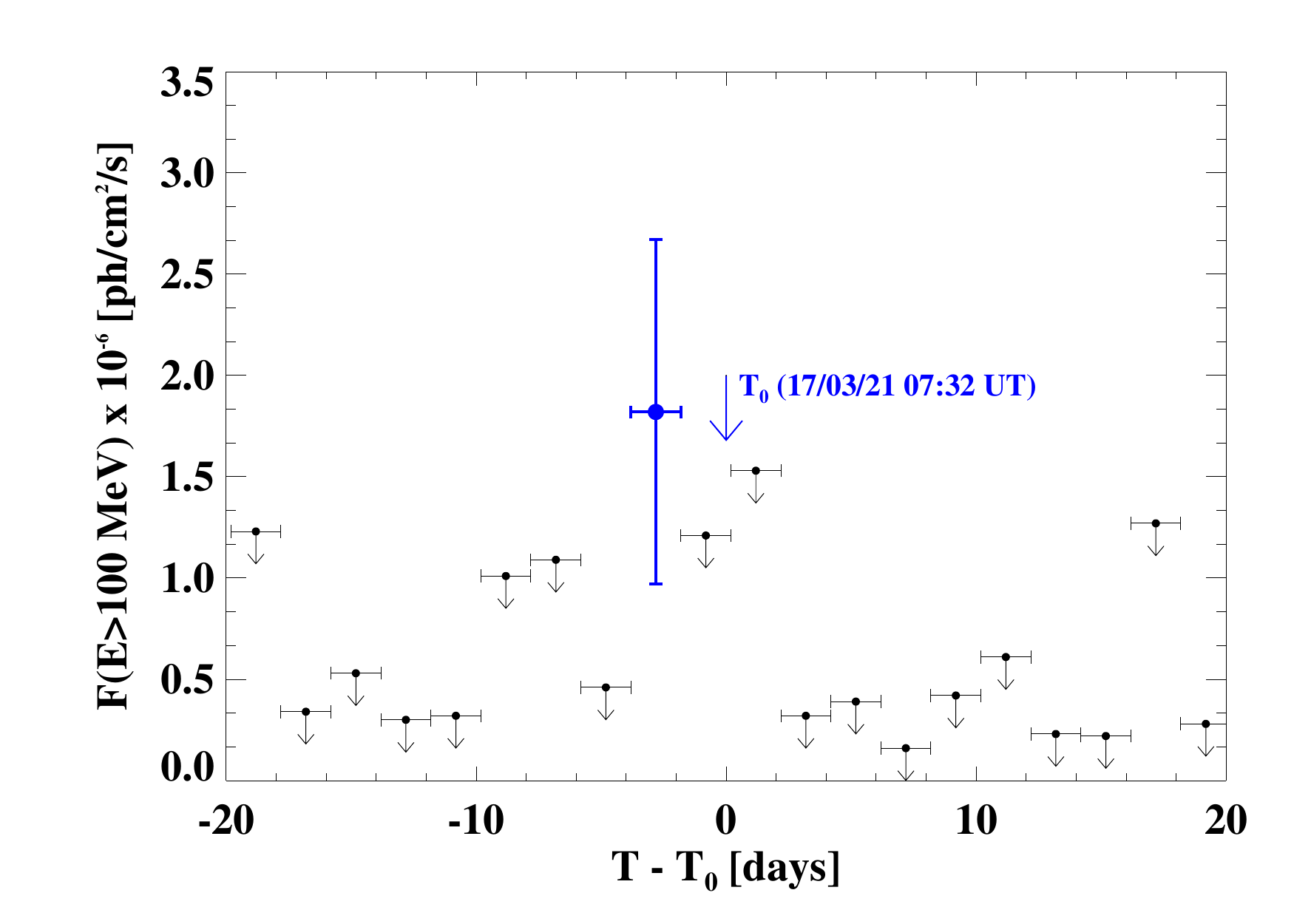}{0.5\textwidth}{}
          \rightfig{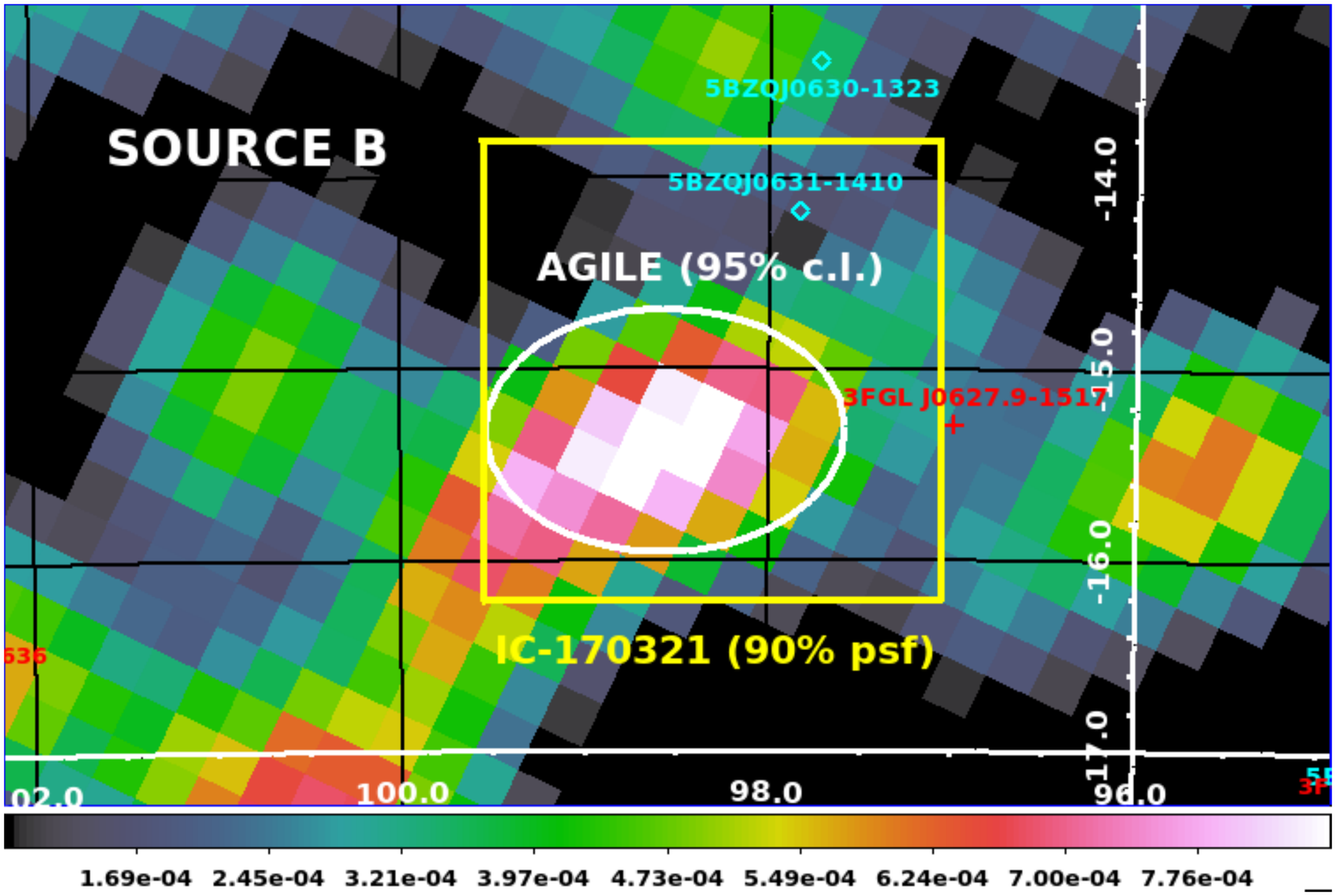}{0.5\textwidth}{}
          }
\vspace{-1.2cm}
\gridline{
          \leftfig{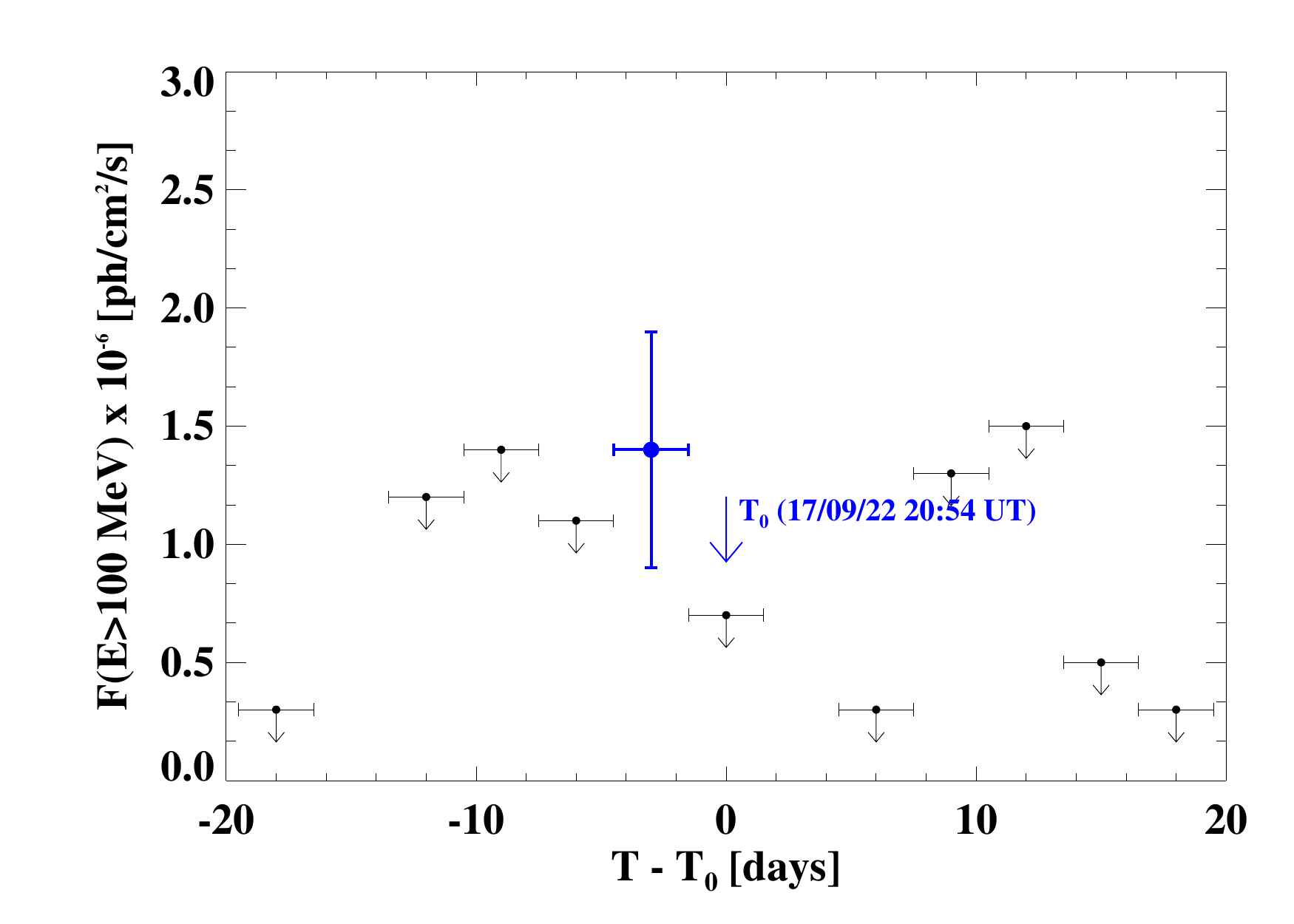}{0.5\textwidth}{}
          \rightfig{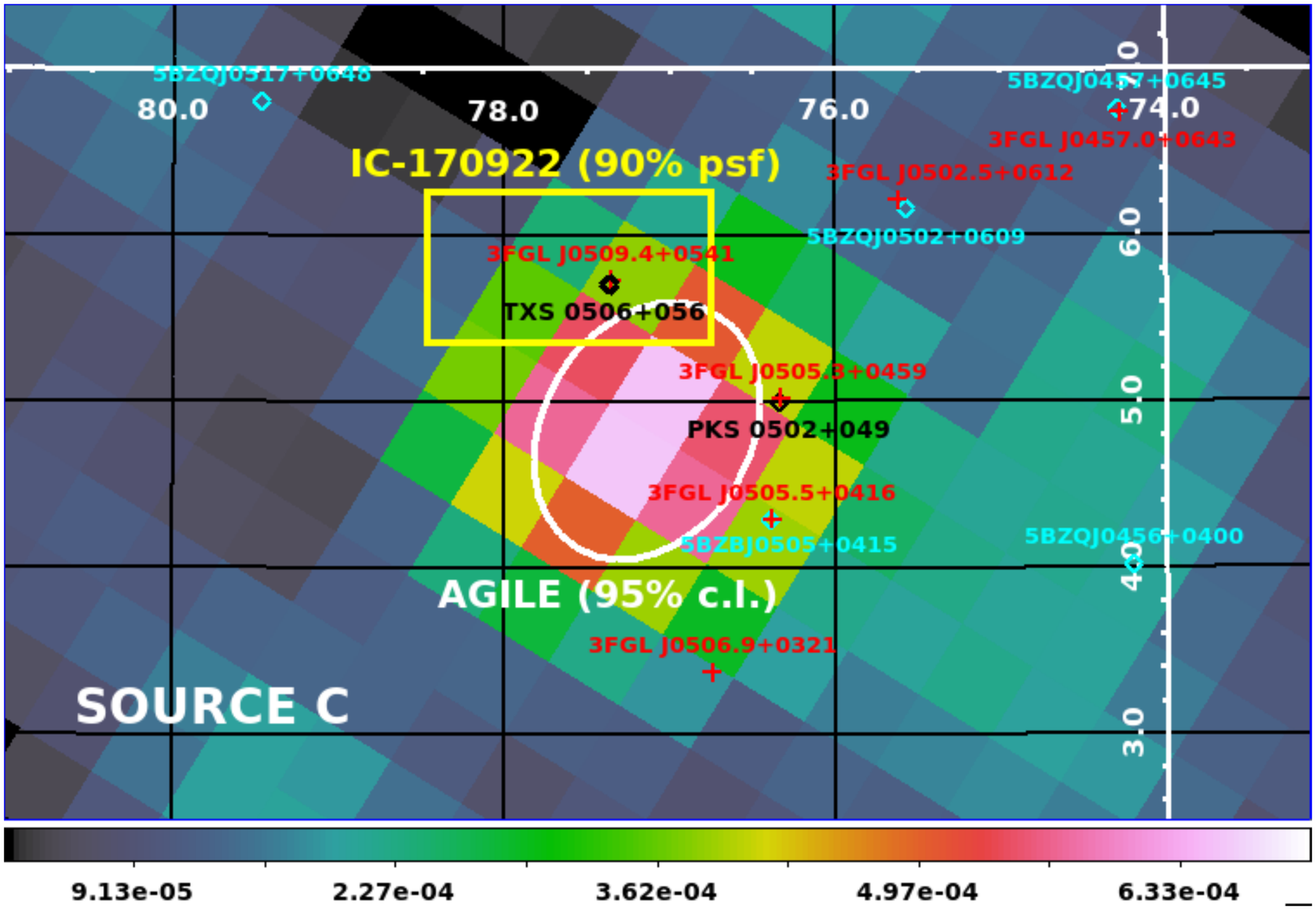}{0.5\textwidth}{}
          }
\vspace{-1.0cm}
\caption{\small %\scriptsize
\replaced{Results of the {\it AGILE}-–GRID standard analysis for 
the three Sources A, B and C of Table~\ref{table_agile_detections}. 
{\it Left panels}: maps in Galactic coordinates of gamma-ray intensity 
above 100~MeV in (ph cm$^{-2}$ s$^{-1}$ sr$^{-1}$) within two days 
from T$_0$. {\it Right panels}: gamma-ray lightcurves above 100~MeV, near 
the IceCube T$_0$, for each of the three sources. The \agile 95\% 
confidence level (c.l.) location contours are shown in white color; 
the IceCube error boxes or circles are shown in black.}
{\edit1{Results of the {\it AGILE}-–GRID standard analysis for the three 
Sources A \edit1{(\it upper panel)}, B \edit1{(\it middle panel) } and 
C \edit1{(\it lower panel) } of Table~\ref{table_agile_detections}. 
{\it Left panels}: gamma-ray lightcurves above 100~MeV around the IceCube T$_0$. 
{\it Right panels}: maps in Equatorial coordinates (J2000) of gamma-ray 
intensity above 100~MeV in (ph cm$^{-2}$ s$^{-1}$ sr$^{-1}$) corresponding 
to the gamma-ray detection before T$_0$ shown on the left. The 
\agile 95\% confidence level (c.l.) location contours are shown in 
white color; the IceCube error boxes or circles are shown in yellow. A systematic 
error of 0.1$^\circ$ should be added to each \agile source determination. 
The positions of the classified AGNs from the BZCAT Catalog~\citep{2015Ap&SS.357...75M} 
and the {\it FERMI}--LAT 3FGL gamma-ray sources~\citep{2015ApJS..218...23A} are shown 
in cyan and red colors, respectively.}}\label{fig1}}
\end{figure*}

All the 3 QL detections are confirmed using the standard 
\agile analysis~\citep{2012A&A...540A..79B}, applying additionally a 
more stringent cut on the Earth albedo contamination\footnote{For 
comparison, the predefined QL maps are generated with a looser Earth albedo 
cut of $80^\circ$.}. \replaced{Figure~\ref{fig1} shows in the {\it left panel}, 
the {\it AGILE}-–GRID intensity maps above 100~MeV, within two days 
from T$_0$, centered at the position of the three sources A, B, C, and 
in the {\it right panel} the corresponding gamma-ray lightcurves 
around T$_0$.}{\edit1{Figure~\ref{fig1} shows: in the {\it left panel}, 
the {\it AGILE}-–GRID gamma-ray lightcurves above 100~MeV around T$_0$ for 
each of the three sources; in the {\it right panel}, the 
gamma-ray intensity maps above 100~MeV corresponding to the 
detection found near T$_0$.}} 

\begin{figure*}[t!]
\gridline{\leftfig{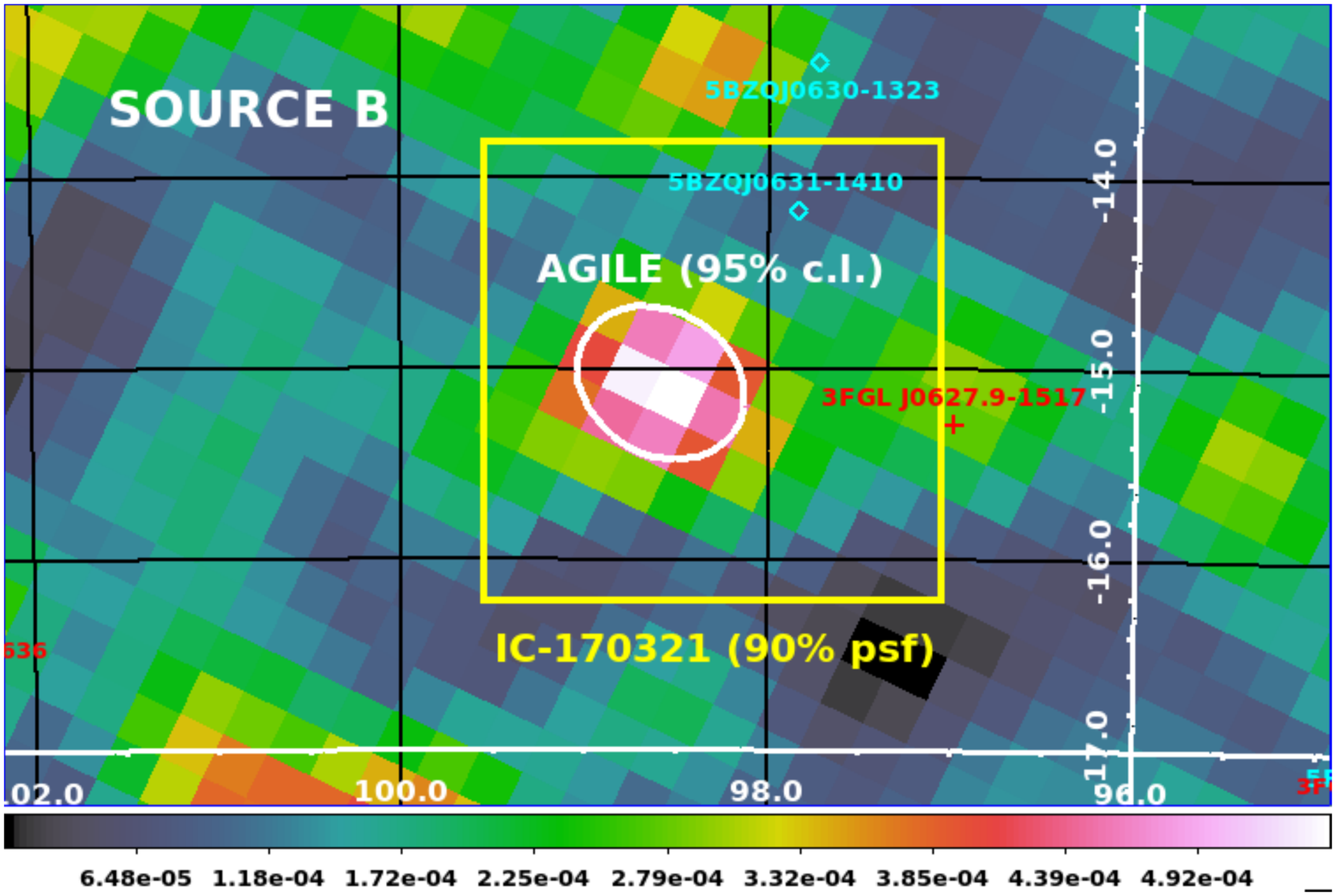}{0.5\textwidth}{}
          \rightfig{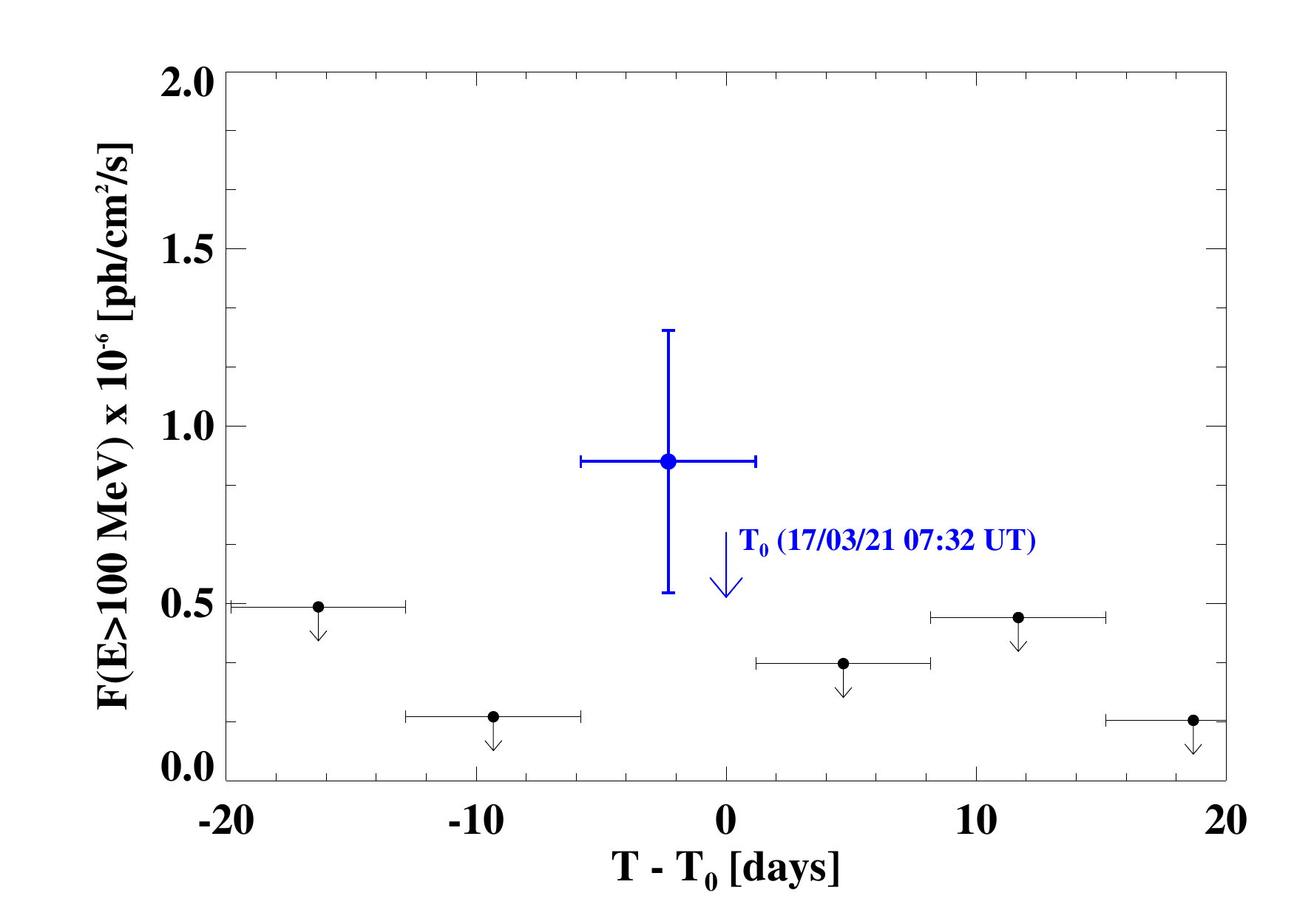}{0.5\textwidth}{}
          }
\vspace{-1.0cm}
\caption{{\it Left panel}: {\it AGILE}–-GRID intensity maps, in 
(ph cm$^{-2}$ s$^{-1}$ sr$^{-1}$) and \replaced{Galactic coordinates}
{\edit1{Equatorial coordinates (J2000)}}, centered at the position 
of the IceCube event IC-170321, over a long integration 
time of 15 days around T$_0$ ((T$_0$ - 6; T$_0$ + 11) days). 
\added{The \agile 95\% c.l. location contour obtained 
with the \agile standard analysis is shown in white; the 
IceCube error box in yellow.} 
{\it Right panel}: {\it AGILE}–-GRID 7-day time bin 
gamma-ray light curve (E$>$100~MeV) around T$_0$, obtained from 
the \agile standard analysis performed at the IC-170321 position.\label{fig3}}
\end{figure*}

The standard \agile data analysis indicates that in all cases the peak 
gamma-ray emission is similarly observed within 1–-2 days from T$_0$. For 
sources B (IC-170321) and C (IC-170922), a weak gamma-ray emission is 
also observed over longer integration time-scales that include T$_0$. 
In particular, for the new Source B an integration of 15 days starting 
from March 15th, 2017 (12:00 UT), shows a detection above 
4$\sigma$ with a flux $F(E>100 MeV) = (4.6 \pm 1.6) \times 10^{-7} 
\,\rm{ph}\,\rm{cm}^{-2}\,\rm{s}^{-1}$. The \agile centroid has Galactic 
coordinates ($l,b$)=(224.59, -10.53) $\pm$ 0.42 (deg) (95\% stat. c.l.) 
$\pm$ 0.1 (deg) (syst.) (R.A., Decl. (J2000)=(98.58, -15.08) (deg)), and 
it is fully compatible with the IceCube centroid (see Fig.~\ref{fig3}). 

\subsection{Post-trial false alarm probability}\label{post_trials}
\edit1{To evaluate the probability that each of these three 
gamma-ray sources \replaced{be}{\edit2{is}} associated to the neutrino events by chance, 
we have firstly evaluated the {\it false alarm rate (FAR)} for an \agile QL detection 
per unit time $\delta t$ and per unit solid angle $\delta \Omega$. 
As the unit time $\delta t$, we have assumed the standard integration time 
of the QL maps ($\delta t=$~2 days). For $\delta \Omega$, we have assumed 
the solid angle subtended by a cone with half-aperture matching 
the standard circular radius of $1.5^\circ$ used in the database 
search \replaced{($1.5^\circ$)}{\edit2{($\delta \Omega \simeq 2.15 \times 10^{-3}$~sr)}}
\footnote{See Appendix~\ref{FAR} for the details of the FAR computation.}.
\replaced{Considering the method already applied in~\cite{2016ApJ...826L...6C}, 
we then estimate the post-trial false alarm probability $P_i$ 
of a random occurrence in {\it space} and {\it time} of a neutrino and a gamma-ray 
transient event separated in time by an interval $\Delta t$ and in space 
by an angular distance $\Delta {\theta}$ as:}
{\edit2{We then estimate the post-trial false alarm probability
$P_i$ of a random occurrence in {\it space} and {\it time} of a neutrino and a gamma-ray
transient event separated \deleted{in time}by an interval $\Delta t$ and\deleted{in space} 
%by an angular distance $\Delta {\theta}$ as~\citep{2016ApJ...826L...6C}:}}}
by the solid angle $\Delta \Omega$ (corresponding to the angular distance $\Delta {\theta}$) 
as~\citep{2016ApJ...826L...6C}:}}}
\begin{equation}
P_i = N_{i} * \rm{FAR}(\geq \sqrt{\it{TS}}) * \Delta {\it{t}} * (1 
+ \rm{ln}(\Delta {\it{T}}/\delta {\it{t}})) * \, \Delta {\Omega} \\
\label{EQN1}
\end{equation}
\edit1{where $N_{i}$ is the number of trials for a symmetric time 
window, $\rm{FAR}(\geq \sqrt{\it TS})$ is the false alarm rate per 2-day map 
and per unit solid angle for \replaced{{\it AGILE}–-GRID gamma-ray detections}
{\edit2{\agile detections}} above a given significance 
$\sqrt{TS}$~\footnote{See \cite{2012A&A...540A..79B} for the definition 
of an \agile detection based on the value of the test statistic $TS$ obtained after 
the application of the \agile multi-source maximum likelihood (ML) algorithm.}, 
$\Delta{t}$ is the absolute time difference between the QL detection centroid 
and T$_0$, and $\Delta{T}$ is the one-sided time interval over which the search 
is done (set beforehand to $\Delta{T}=$~4 days).
\replaced{Given that we have assumed a spatial coincidence whenever the 
centroids of the \agile/IceCube detections are within an angular distance 
$\Delta {\theta}=1.5^\circ$, in our case $\Delta {\Omega}$ is equal to the unit solid angle assumed 
as the spatial region of interest, that is $\Delta{\Omega}\equiv\delta{\Omega}$.}
{\edit2{We have assumed a spatial coincidence whenever the centroids 
of the \agile/IceCube detections are within an angular distance $\Delta {\theta}=1.5^\circ$, 
so that in our case $\Delta \Omega \equiv \delta \Omega$.}}}

\deleted{Since the astrophysics and the timescales of the phenomena related to the 
emission of these extremely high-energy neutrinos and their likely correlated 
gamma-ray emission are still uncertain, and based on the typical \agile sensitivity 
to a transient gamma-ray source, we adopted a search time window of interest 
of \edit1{plus/minus} 4 days (== 2 QL maps) around T$_0$. Then, $\Delta{t} = 4$ days. 
Moreover,} \added{\edit1{Since}} the gamma-ray detection strategy we adopted is fully 
automated\footnote{The start and stop of the 2-day time integration has 
been defined {\it a-priori} since the start of the spinning observation mode.}, 
and there is no refined analysis around T$_0$\replaced{: so}{,} the trials 
factor $N_{i}$ takes into account only the choice of the symmetric 
window around T$_0$ and is thus equal to 2.

The last two columns of Table~\ref{table_agile_detections} show, 
respectively, the FAR (per 2-day map and per unit solid angle) and the corresponding 
{\it post-trial} false alarm probability $P_i$ of a random coincidence 
with the IceCube neutrinos for each of the 3 QL detections. For each 
\agile source, the {\it post-trial} chance correlation is of the order of $10^{-3}$. 

Given this basic information, we then proceed to calculate the {\it joint} 
post-trial probability to observe 3 gamma-ray sources out of 10 neutrino 
alerts over the period of the active IceCube alert system, as:
$$ P_{joint} (\rm{post-trial}) = 1 - (1 - {\it P_A} * {\it P_B} * {\it P_C})^{\it N} $$
where the number of global trials $N$ is given by the 
product of two contributions: the total number of IceCube HESE/EHE events considered 
(equal to 10), and the number (equal to 3) of optimizations of the search 
radius of the gamma-ray error boxes. We therefore determine the joint 
{\it post-trial chance probability} to be:
$$ P_{joint} (\rm{post-trial}) = 1.7 \times 10^{-6} $$
which corresponds to a number of Gaussian equivalent one-sided standard 
deviations of approximately 4.7$\sigma$.

Alternatively, assuming an average {\it post-trial} false alarm probability 
$p = 4.0 \times 10^{-3}$ for a typical gamma-ray source, we can use a 
binomial probability distribution to estimate the probability to 
observe 3 \agile gamma-ray counterparts out of 10 IceCube events in the whole sky. 
That results in a probability of the order of $7.5 \times 10^{-6}$ 
(one-sided 4.3$\sigma$).

\vspace{1cm}

\section{Possible e.m. counterparts to the IceCube events and the sources 
A, B and C detected by AGILE}\label{counterparts}
\vspace{0.3cm}
\subsubsection{\agile Source A/IC-160731 event}
\begin{figure*}[t!]
\gridline{\leftfig{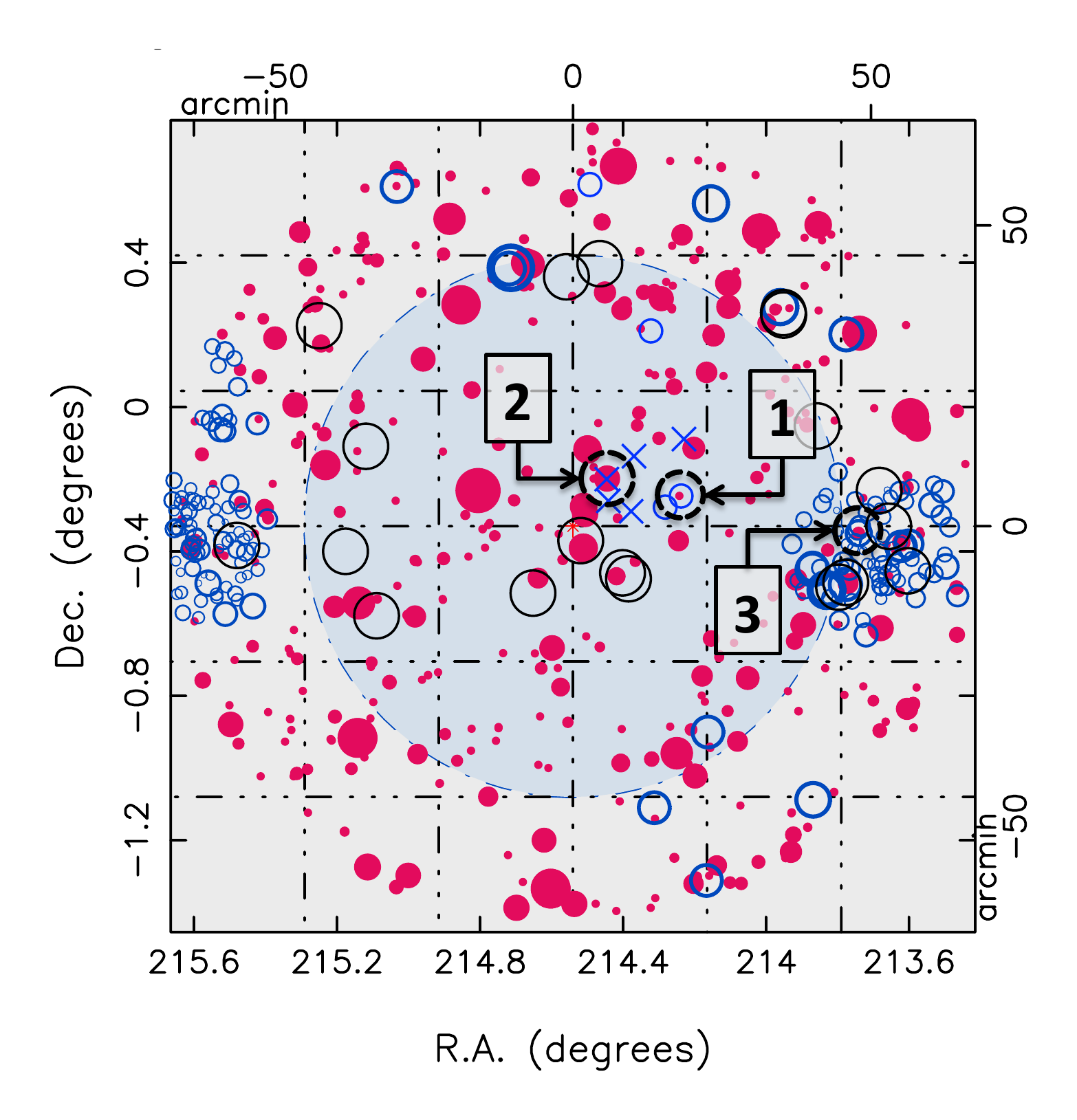}{0.4\textwidth}{}
          \rightfig{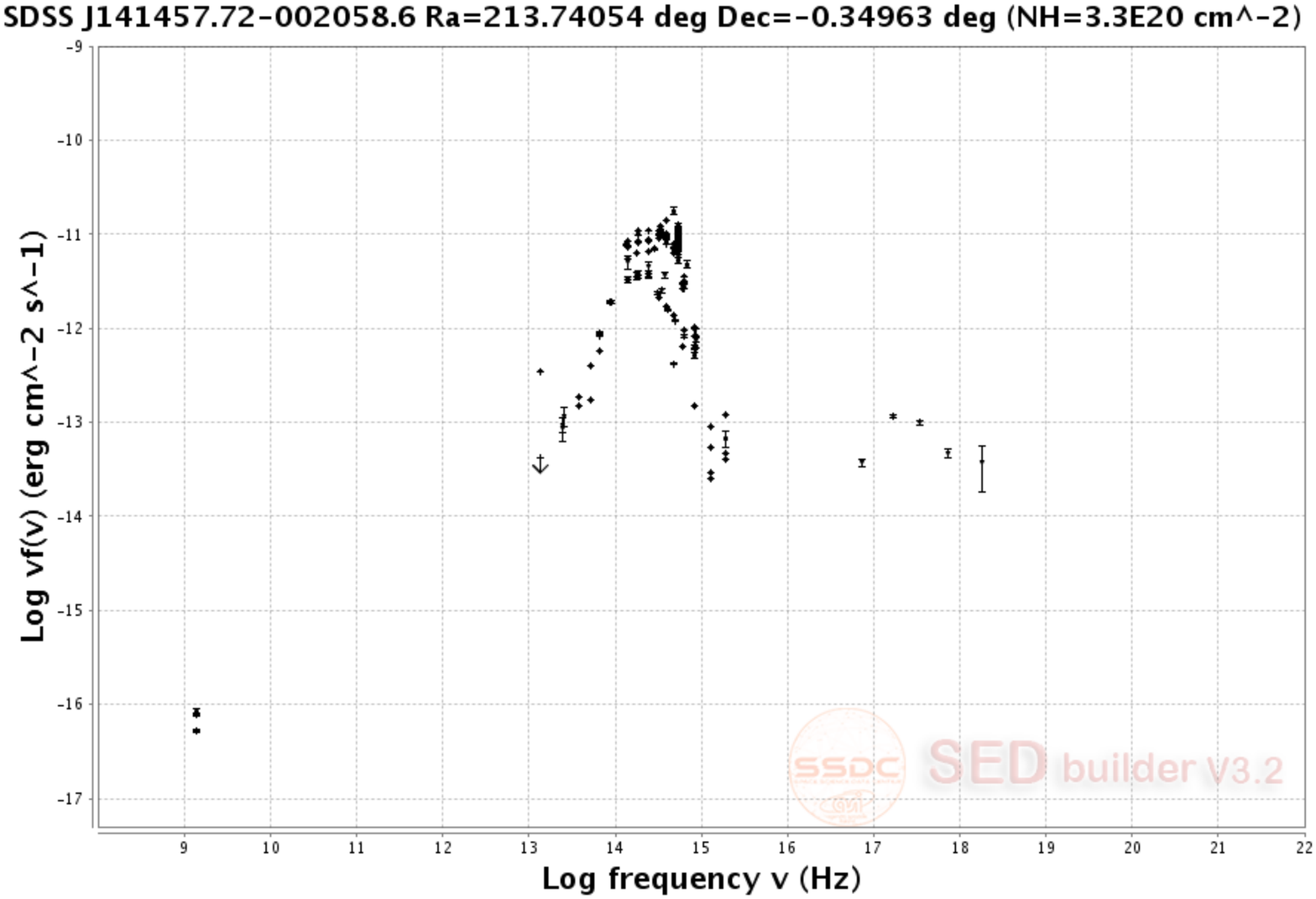}{0.57\textwidth}{}
          }
\vspace{-1.0cm}
\caption{{\it Left panel}: R.A.--Decl. sky map (J2000) obtained using the SSDC 
{\it SkyExplorer} tool\textsuperscript{\ref{skyexplorer}} and the ASI {\it Open Universe} 
tool\textsuperscript{\ref{openuniv}} showing known radio (red filled circles) 
and X-ray (open blue circles) sources within 60 arcmin 
from the IC-160731 centroid (R.A., Decl. (J2000) = (214.544, -0.3347) deg). 
\added{\edit1{The blueish circular area represents the position uncertainty (90\% c.r.) quoted 
by the IceCube Coll. for the IC-160731 event (see Table~\ref{ICECUBE_ALERTS}).}} 
The map also covers most of the 95\% c.l. contour of the \agile Source A (already 
known as AGL J1418 +0008), centered at R.A., Decl. (J2000)=(214.61, 0.13 deg). 
Black open circles are all known Galaxy Clusters from existing catalogues. 
Source intensities are related to the circle diameters. Source labeled as 
$1$ was already studied in~\citep{2017ApJ...846..121L}. Blue crosses indicate 
the positions of 5 uncatalogued X-ray sources detected during a 
dedicated Swift observation of source $1$~\citep{2017ApJ...846..121L}. Source 
labeled as $3$ is a possible HBL blazar candidate. {\it Right panel}: SED of the 
HBL/HSP blazar candidate labeled as $3$ in the figure on the left, obtained 
with archival data from radio to X-rays available at the SSDC. \added{\edit1{Recently, 
this object have been included in the 3rd edition of the HSP blazar catalog 
with the name of 3HSP J141457.7-002058~\citep{3HSP}.}}\label{fig4}}
\end{figure*}

The first IceCube HESE/EHE event, compatible and temporally close 
to an automatic \agile QL detection, occurred on July 31, 2016 (T$_0$ = MJD 57600.079). 
The event and the possible \agile gamma-ray counterpart (AGL J1418 +0008) were 
extensively studied in \cite{2017ApJ...846..121L}. The e.m. follow-up of the event 
did not reveal any transient sources within the IceCube error circle. Using the online 
SSDC {\it SkyExplorer} tool\footnote{https://tools.ssdc.asi.it\label{skyexplorer}} 
and the ASI {\it Open Universe} web portal\footnote{http://www.openuniverse.asi.it\label{openuniv}}, 
in this work we have performed a new search for possible known e.m. counterparts 
within the common \agile/IC-170321 confidence error regions. 
Figure~\ref{fig4}, {\it left panel}, shows the result of a 
query for cataloged radio, X-ray and gamma-ray sources within 60 arcmin from the 
IceCube centroid, placed at R.A., Decl. (J2000)=(214.544, -0.3347 deg). The 60 arcmin 
search radius encompasses the whole IC-160731 error circle and also covers most of the 
95\% c.l. error circle of the \agile Source A detection (see Fig.~\ref{fig1}, 
{\it upper panel}).

The sky region within the gamma-ray and neutrino error regions does not 
show any obvious e.m. counterpart, in particular, neither known gamma-ray 
sources nor known AGN blazars appear within the search radius chosen for the query. 
The X-ray source 1RXS J141658.0−001449 (labeled as $1$ in Fig.~\ref{fig4}) 
was suggested in \cite{2017ApJ...846..121L} as a potential 
high-peaked BL Lac (HBL) AGN blazar. Nevertheless, a dedicated Swift-XRT 
observation taken some months after the neutrino event time T$_0$, did not confirm 
any steady X-ray emission from this position, thus the former hypothesis could not 
be confirmed. 

Five uncatalogued X-ray sources were detected during the previous Swift target 
of opportunity (ToO)~\citep{2017ApJ...846..121L}: their positions are 
indicated by the blue crosses in Fig.~\ref{fig4}. 
One of them (source labeled as $2$) is positionally 
consistent with the radio source NVSS J141746-001151 and the object 
SDSS J141746.65-001149.8, which is actually catalogued as star. 

Interestingly this region shows the presence of several galaxy clusters 
(indicated by the black circles in Fig.~\ref{fig4}), which could 
host a possible AGN or a different class of powerful active sources that 
can be the origin of the IceCube neutrino and the gamma-ray transient emission 
detected by AGILE. In particular, based on its radio/X-ray positional association 
and flux intensity, one of the most interesting neutrino source candidates within 
this sky region is the source labeled as $3$ in Fig.~\ref{fig4} 
(R.A., Decl. (J2000)=213.74038, -0.34967 deg). The radio and X-ray emissions 
are positionally consistent with the elliptical galaxy SDSS J141457.72-002058.6, 
whose broadband spectral properties resemble those ones typical of a 
\edit1{high synchrotron peaked (HSP)} blazar (see Fig.~\ref{fig4}, 
{\it right panel})\footnote{\added{\edit1{Indeed, this object have been recently included in the 
3rd edition of the extreme and HSP blazars catalog as 3HSP J141457.7-002058~\citep{3HSP}.}}}.
\vspace{0.3cm}
\subsubsection{\agile Source B/IC-170321 event}\label{ic170321counterparts}
The second IceCube HESE/EHE event, compatible and temporally close 
to an automatic \agile QL detection, occurred on March 21, 2017 
(T$_0$ = MJD 57833.314). The ML significance of the QL detection is slightly 
lower than the other ones but it is again confirmed through the standard 
\agile analysis using a longer integration window around T$_0$, applying 
additionally a more stringent cut on the Earth albedo contamination.

The e.m. follow-up of the event did not reveal any transient source 
within the IceCube error box: in the HE gamma-ray band, {\it FERMI}--LAT placed a 
95\% C.L. upper limit (u.l.) above 100~MeV for point-like emission 
at the IceCube position over different time intervals near and before T$_0$, with 
the most stringent found to be $5.5 \times 10^{-8}$ ph cm$^{-2}$ s$^{-1}$ 
in one week of exposure prior to T$_0$~\citep{2017GCN.20971....1B}.

\begin{figure}[t!]
\centering
\includegraphics[scale=0.60]{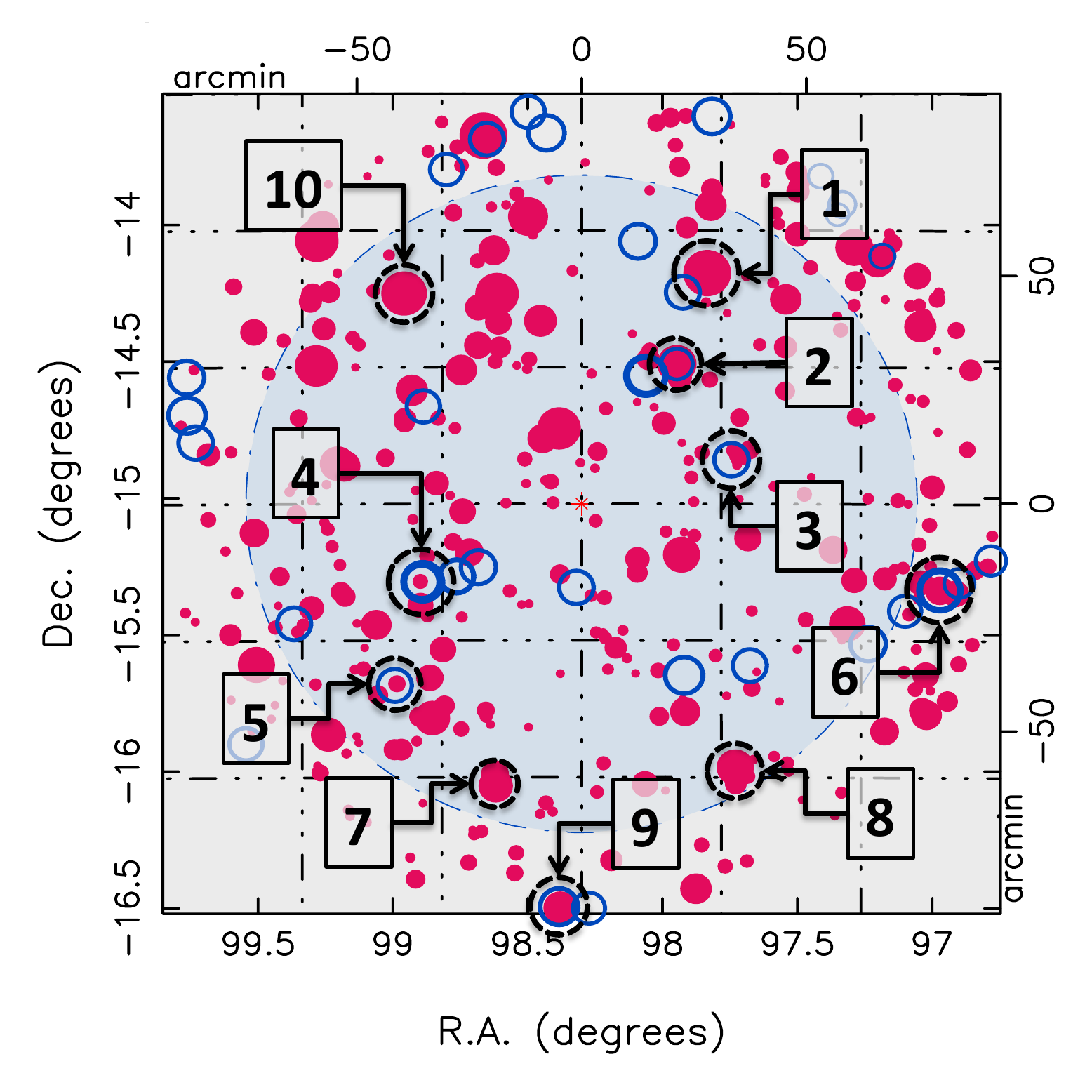}
\caption{R.A.-Decl. sky map (J2000) obtained using the SSDC {\it SkyExplorer} tool 
and the ASI {\it Open Universe} tool showing known radio, X-ray and 
gamma-ray sources within 90 arcmin from the IC-170321 
centroid (R.A.-Decl. (J2000): 98.3, -15.02 (deg)). 
\added{\edit1{The blueish circular area represents the uncertainty (90\% PSF) 
on the reconstructed neutrino arrival direction quoted by the IceCube Coll. for 
the IC-170321 event (see Table~\ref{ICECUBE_ALERTS}).}}
The map also covers the whole 95\% c.l. error circle of the \agile Source B, 
centered at R.A., Decl. (J2000)=(98.58,  -15.08 deg). Radio (red filled circles) 
plus X-ray (open blue circles) sources map from existing catalogues. \replaced{Map showing 
blazar candidates from the same sky area (details of candidates, labeled as $1$ to $10$, are 
reported in Table~\ref{ic170321_candidates}). Filled circles size is proportional 
to radio flux density, open circles size is proportional to X-ray flux. 
Orange symbols: HSP/HBL blazar candidates; Cyan: intermediate BL Lac (IBL) 
candidates; Red: radio-sources with no X-ray counterparts; 
Black: unknown Blazar types and/or Galaxy clusters.}{\edit1{Possible common e.m. 
candidate counterparts are enclosed by black dashed circles and labeled 
from $1$ to $10$. Details of each candidate is reported in 
Table~\ref{ic170321_candidates}.}}
\label{fig5}}
\end{figure}

In the hard X-ray/gamma-ray band, INTEGRAL and Konus-Wind 
reported upper limits on burst-type emission over short time periods 
around T$_0$~\citep{2017GCN.20937....1S, 2017GCN.20973....1S}. 
The Swift-XRT follow-up, with a 7-tile mosaic covering only 21\% 
of the 90\% error box on the refined IceCube localization, detected only one 
known X-ray source (1SXPS J063214.5-143300) at a flux level consistent 
with the cataloged value~\citep{GCN20964}. Archival data from 
Swift-BAT\footnote{https://swift.gsfc.nasa.gov/results/transients/index.html} 
did not show any transient hard-X-ray emission at this position. No optical 
follow-up was reported for the event. We explored the All-Sky Automated 
Survey for SuperNovae (ASAS-SN) transient web 
page\footnote{http://www.astronomy.ohio-state.edu/asassn/transients.html} and the Palomar 
Transient Factory (PTF) catalog\footnote{http://irsa.ipac.caltech.edu/Missions/ptf.html} 
but did not find any transient optical emission within $1^\circ$ from the IceCube centroid.

\begin{deluxetable*}{lcccccc}
\tablecaption{\replaced{Possible e.m. candidate counterparts found within 
the common \agile Source B error region and the IceCube error box 
of the IC-170321 neutrino event.}{Possible e.m. candidate counterparts 
for IC-170321 and the \agile Source B detected in the days around T$_0$.}\\\label{ic170321_candidates}}
\tablewidth{0pt}
\tabletypesize{\scriptsize}
\setlength{\tabcolsep}{0.02in}
\tablehead{
\\ \colhead{ID} & \colhead{Catalog name} & \colhead{R.A. (J2000)} &
\colhead{Decl. (J2000)} & \colhead{Other association} 
& \colhead{Source class} & \colhead{Distance from IC-170321 centroid} \\
\colhead{} & \colhead{} & \colhead{(deg)} & \colhead{(deg)} &
\colhead{} & \colhead{} & \colhead{(arcmin)}
}
\startdata
1 & 5BZQ J0631-1410       & 97.83429 & -14.1755  & CRATES J063119-141030 & FSRQ & 58 \\ \hline
2 & CRATES J063148-143042 & 97.94638 & -14.50844 & -- & Possible IBL & 37 \\ \hline
3 & PSZ2 G224.01-11.14    & 97.75250 & -14.83520 & -- & Cluster of Galaxies & 34 \\ \hline
4 & NVSS J063535-151813   & 98.89838 & -15.30361 & 1RXS J063533.5-151817 & Possible HBL & 39 \\ \hline
5 & NVSS J063556-154038   & 98.98450 & -15.67736 & 1RXS J063558.2-15410 & Possible HBL & 56 \\ \hline
6 & 3FGL J0627.9-1517     & 96.9853  & -15.29782 & WHSP J062753.2-151956 & HSP BL Lac & 79 \\ \hline
7 & CRATES J063428-160239 & 98.6191  & -16.0519  & -- & Flat spectrum radio source & 65 \\ \hline
8 & CRATES J063053-155929 & 97.7306  & -15.9829  & -- & Flat spectrum radio source & 67 \\ \hline
9 & CRATES J063329-163020 & 98.38046 & -16.5201  & -- & Possible HBL & 89 \\ \hline
10& PMN J0635-1415        & 98.95842 & -14.25011 & -- & Flat spectrum radio source & 60 \\ \hline
\enddata
\end{deluxetable*}

Using the online SSDC {\it SkyExplorer} tool\textsuperscript{\ref{skyexplorer}} and 
the ASI {\it Open Universe} web portal\textsuperscript{\ref{openuniv}}, we 
searched also for this event a possible \added{\edit1{common}} e.m. counterpart 
\replaced{within the common \agile/IC-170321 confidence error 
regions.}{\edit1{for the IC-170321 neutrino and the \agile Source B.}} 
Figure~\ref{fig5}\deleted{, {\it left panel},} shows the result of a 
query for cataloged radio, X-ray and gamma-ray sources 
within 90 arcmin from the IceCube centroid\added{\edit1{, which fully contains 
the \agile Source B error circle}}. Labels from $1$ to $10$ in 
Figure~\ref{fig5}\deleted{, {\it right panel},} indicate the most interesting 
neutrino/gamma emitter candidates found in the search, based on 
their radio/X-ray positional association and flux intensity. 
Among them, we found two flat spectrum radio quasars (FSRQ), one 3FGL source, 
3FGL J0627.9-1517, and three possible blazars of the HBL sub-class. 
Details of each of the 10 sources are reported in Table~\ref{ic170321_candidates}.

Assuming the HBL sub-class of blazars as one of the most promising neutrino 
emitter candidates~\citep{2016MNRAS.457.3582P, 2017MNRAS.468..597R}, one of the 
most interesting source within the {\it SkyExplorer} search radius appears 
to be the 3FGL J0627.9-1517 source (\#6 in Fig.~\ref{fig5}\deleted{, {\it right panel}}), 
which has been recently classified as the high synchrotron peaked (HSP) blazar 
2WHSP J062753.2-151956~\citep{2017A&A...598A..17C}. 
The steady average 3FGL flux above 100~MeV for this source is 
below $2 \times 10^{-8} \, \rm{ph} \, \rm{cm}^{-2} \, \rm{s}^{-1}$. 
The gamma-ray light curve above 1~GeV produced with the {\it FERMI}--LAT online 
data analysis tool available at SSDC\footnote{https://tools.asdc.asi.it/?\&searchtype=fermi\label{fermiaphot}}, 
with a 7-day binning, did not show any relevant 
activity in the six months around the neutrino event T$_0$, except 
for one little peak found some days after which it was not 
confirmed by a further analysis made with the official 
{\it FERMI} Science Tools (v10r0p5)\footnote{https://fermi.gsfc.nasa.gov/ssc/data/analysis/software/}. 

\vspace{0.3cm}
\subsubsection{\agile Source C/IC-170922 event}
The first \agile detection of a gamma-ray counterpart 
above 100~MeV consistent with the position of the neutrino 
event IC-170922 was firstly reported in \cite{2017ATel10801....1L}. 
Again, the detection initially appeared as result of the automatic QL daily 
processing, and was confirmed afterwards using the standard \agile analysis. The e.m 
follow-up triggered by the GCN Notice and the GCN Circular 
announcing the identification of an EHE neutrino event by IceCube~\citep{GCN21916} 
allowed the identification of the blazar BL Lac TXS 0506+056 
(also known as 5BZB J0509+0541~\citep{2015Ap&SS.357...75M}) as the likely 
counterpart of the IceCube event~\citep{2018Sci...361.1378I}.

\begin{figure*}[t!]
\gridline{
  \leftfig{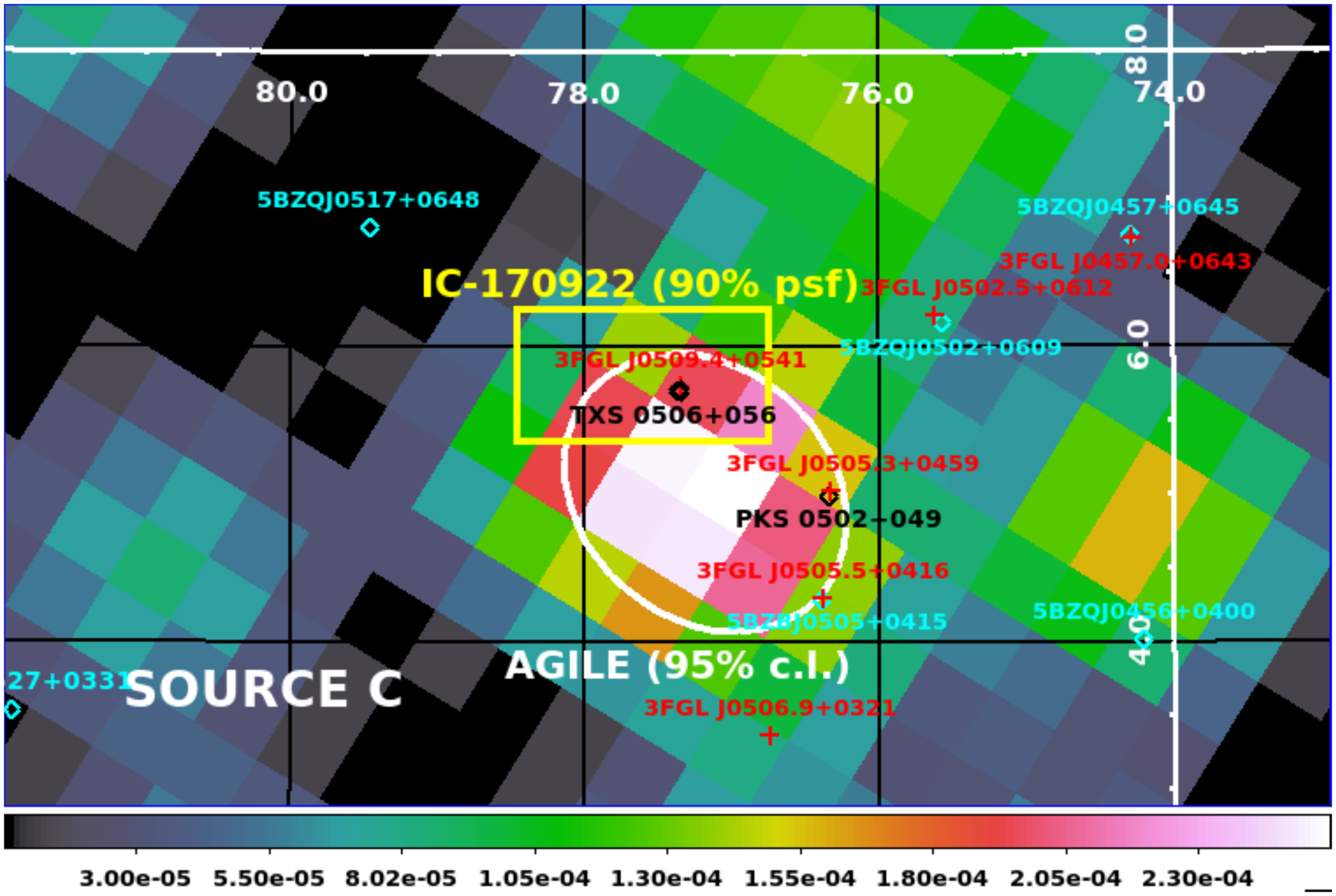}{0.5\textwidth}{}
  \rightfig{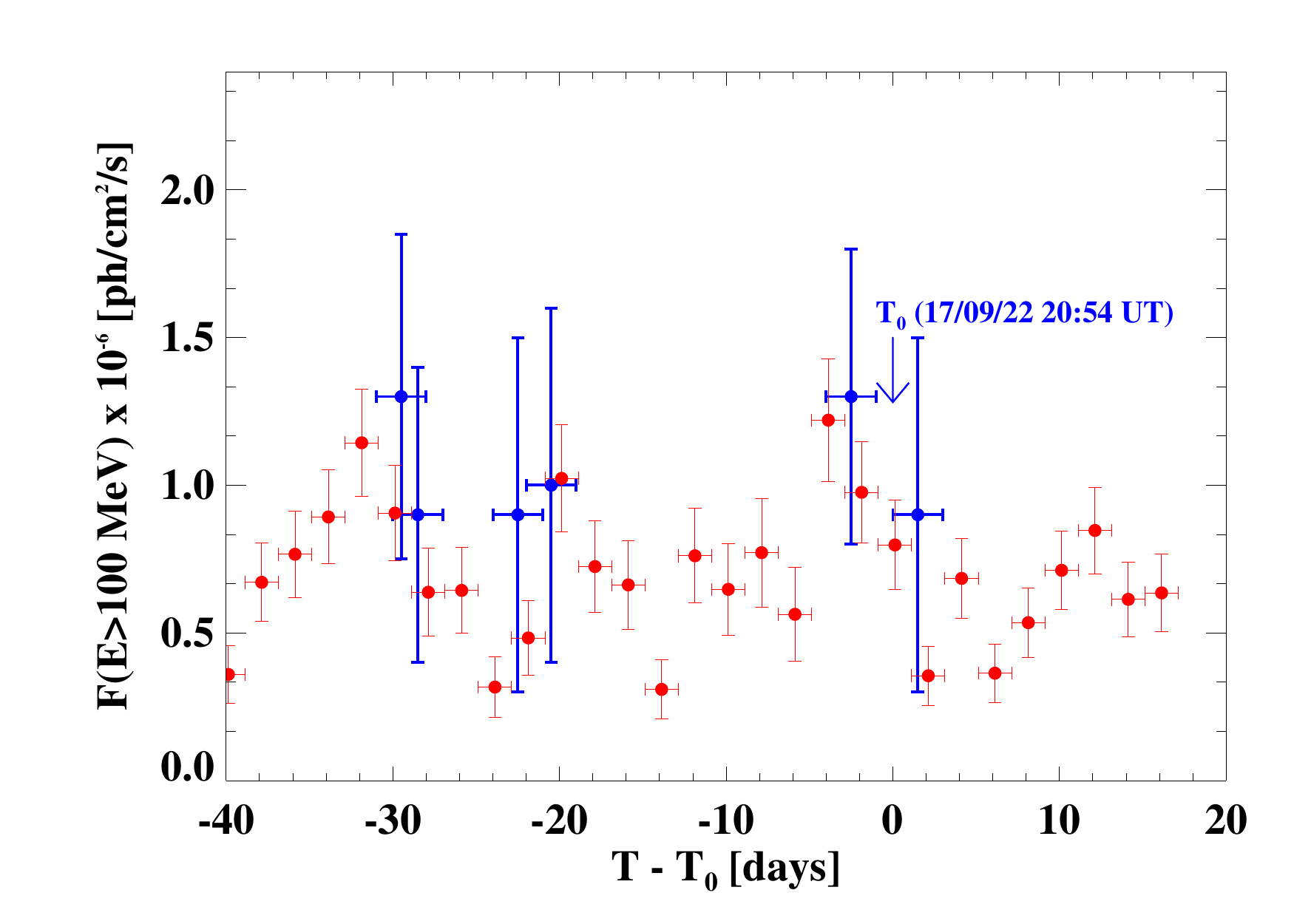}{0.5\textwidth}{}
          }
\vspace{-1.0cm}
\caption{{\it Left panel}: {\it AGILE}–-GRID intensity map above 400~MeV, 
in (ph cm$^{-2}$ s$^{-1}$ sr$^{-1}$) and Equatorial coordinates (J2000), around the 
region of the TXS 0506+059 source, over the period (T$_0$-4; T$_0$-1) days. 
The \agile 95\% c.l. contour obtained with the \agile standard analysis is 
shown in white color; the IceCube error box~\citep{GCN21916} is shown in \replaced{black}{\edit1{yellow}}. 
The positions of the classified AGNs from the BZCAT Catalog~\citep{2015Ap&SS.357...75M} 
and the FERMI-LAT 3FGL gamma-ray source catalog~\citep{2015ApJS..218...23A} are 
shown in cyan and red colors, respectively. {\it Right panel}: 
\agile gamma-ray lightcurve above 100~MeV on the TXS 0506+059 position, in the days 
before and after the T$_0$ of the IceCube IC-170922 event. In red, the corresponding 
{\it FERMI}--LAT lightcurve from the same position, obtained with the 
{\it FERMI}--LAT online analysis tool publicly available 
at the SSDC\textsuperscript{\ref{fermiaphot}}.
\label{fig6}}
\end{figure*}

Using GRID data with energies above 400~MeV in a time interval 
of three days close to the neutrino event T$_0$, we obtained 
a better positional agreement of the \agile detection with 
the TXS 0506+059 source, contained within the IC-170922 error box (see 
Fig.~\ref{fig6}, {\it left panel}), which thus confirms 
the gamma-ray activity observed from the source during this 
period~\citep{2017ATel10791....1T, 2018Sci...361.1378I}.

As reported in~\citep{2018Sci...361.1378I}, the source has been active 
in gamma-rays since several months before Sept. 2017. Figure~\ref{fig6}, {\it right panel}, 
shows the \agile gamma-ray lightcurve above 100~MeV from the 
beginning of August till the end of Sept. 2017, estimated on the TXS 0506+056 position. 
Superimposed is the corresponding {\it FERMI}--LAT curve (red points) obtained with 
the public analysis tool available at SSDC\textsuperscript{\ref{fermiaphot}}, 
which shows a good agreement with the flaring activity detected by 
\agile\footnote{We notice that the {\it FERMI}--LAT fluxes estimated with the online tool 
can be overestimated up to a factor of 2.}.

A recent IceCube paper claimed a second excess of VHE neutrinos observed from the direction 
of TXS 0506+056 in the period Sept. 2014 -– beginning of 2015~\citep{2018Sci...361..147I}. 
The analysis of the {\it AGILE}-–GRID data over this period 
around the TXS 0506+059 position shows a strong gamma-ray 
contribution from the near FSRQ source PKS 0502+049 ($1.2^\circ$ away), 
which was in a high flaring state at that epoch~\citep{2014ATel.6425....1O, 
2014ATel.6457....1L}. Using {\it FERMI}--LAT data, \cite{2018MNRAS.480..192P} show that 
the gamma-ray emission from the TXS is particularly hard compared to the 
softer emission from the FSRQ, and becomes predominant only selecting 
gamma-rays above the GeV. Indeed, our analysis also shows that the contribution 
from PKS 0502+049 above a few GeV becomes negligible but, 
due to the limited \agile gamma-ray sensitivity above 1~GeV, we can 
\replaced{place only}{\edit1{set a flux UL}} $< 3.8 \times 10^{-8} \,\rm{ph}\,\rm{cm}^{-2}\,\rm{s}^{-1}$
for E$>1$~GeV (for a 95\% c.l.) over the emission from the 
TXS 0506+056 during this period.

\section{Discussion and conclusions}\label{discussion}
We reported the results of the \agile \gray observations of the 
error regions of 10 IceCube HESE/EHE neutrino events announced since 
April 2016 through the GCN/AMON system.

Mining the database of automated {\it AGILE}--GRID {\it QL} 
detections, determined on predefined 2-day integration maps, 
we found three significant \gray detections above 100~MeV within 
1.5 degrees from the IceCube best-fit centroids, and within 
two days from the neutrino event time T$_0$. The \agile automatic 
detections, each of them with a significance at the level of 4$\sigma$ 
estimated using the \agile ML algorithm, are compatible in 
position and time with the three following IceCube events: IC-160731, 
IC-170321, and IC-170922, all of them classified as EHE.

We dubbed the three \agile sources as A, B, and C. The global 
post-trial probability found in our study for the \agile/IceCube 
chance correlation for 3 out of 10 events is quite low 
(around 4.7 Gaussian standard deviations), and significantly 
hints towards an astrophysical connection between the \agile detections and 
the observed IceCube single-track events.

A direct correlation between gamma-rays and neutrinos from astrophysical 
sources is expected whenever hadronic emission mechanisms are at work. 
In a hadronic source scenario, we do expect comparable values of the 
gamma-ray/neutrino observed luminosities~\citep{1995PhR...258..173G}. 
We thus estimate the \agile gamma-ray luminosities for each of the 
three sources A, B, and C, and compare them with the corresponding 
neutrino luminosities, assuming a typical timescale of 6 months 
for neutrino production as a product of an underlying hadron acceleration 
and interaction~\citep{2018Sci...361.1378I}. Being the A and B sources 
unidentified, we assume two values for their possible distance: 10 kpc 
(typical of a Galactic object), and redshift z=1 (for an extragalactic object). 
For Source C, we make use of the TXS 0506+056 redshift, z=(0.3365 $\pm$ 0.0010), 
recently estimated by~\citep{2018ApJ...854L..32P}. For the calculation of the 
neutrino luminosities, we adopt the muon neutrino fluence value 
of $2.8 \times 10^{-3}$ erg cm$^{-2}$ estimated in \cite{2018Sci...361.1378I}, 
for which we would expect to detect one high-energy neutrino event 
with IceCube over its entire lifetime\footnote{A power-law neutrino spectrum 
is assumed in this estimation with an index equal to -2 between 200 TeV 
and 7.5 PeV~\citep{2018Sci...361.1378I}.}.

\begin{table}[t!]
\begin{center}
\caption{Gamma-ray and neutrino isotropic luminosities for the three 
sources detected by \agile possibly related to three IceCube HESE/EHE neutrinos. 
Gamma-ray luminosities are estimated over a time interval of about $\pm 1$ week 
around T$_0$; for neutrino luminosities, an active source period of 6 months is assumed. 
For Sources A and B, two possible values of distance are considered: D=10 kpc, 
for a typical Galactic object, and redshift z=1, for an \edit1{extragalactic} 
one\tablenotemark{a}. For Source C, \added{\edit1{only}} the estimated redshift 
of TXS 0506+059 ($z$=0.3365)~\citep{2018ApJ...854L..32P} has been used 
for the calculation.}
\label{table_luminosities}
\vskip .3cm
%\small
\scriptsize
%\begin{minipage}{10cm}
\begin{tabular}{l|c|c|c|c|c|c|c|c}
\hline\hline
\agile & IceCube & $\nu F_{\gamma}(\nu)$ & \multicolumn{2}{c|}{\multirow{2}{*}{$D=$10 kpc}} 
& \multicolumn{2}{c|}{\multirow{2}{*}{$z=1$}} & \multicolumn{2}{c}{\multirow{2}{*}{$z=0.3365$}} \\
source & event & $(\rm erg\, cm^{-2}\, s^{-1})$ & \multicolumn{2}{c|}{} & \multicolumn{2}{c|}{} 
& \multicolumn{2}{c}{} \\
\hline
 & & & $L_{\gamma}$ & $L_{\nu}$ & $L_{\gamma}$ & $L_{\nu}$ & $L_{\gamma}$ & $L_{\nu}$ \\ 
 & & & $(\rm erg\, s^{-1})$ & $(\rm erg\, s^{-1})$ & $(\rm erg\, s^{-1})$ & $(\rm erg\, s^{-1})$ 
& $(\rm erg\, s^{-1})$ & $(\rm erg\, s^{-1})$ \\ \hline
A & IC-160731 & $6.9 \times 10^{-11}$ & $8.2 \times 10^{35}$ & $2.2 \times 10^{36}$ 
& $2.6 \times 10^{46}$ & $6.8 \times 10^{46}$ & -- & -- \\ \hline
B & IC-170321 & $7.5 \times 10^{-11}$ & $9.0 \times 10^{35}$ & $2.2 \times 10^{36}$ 
& $2.8 \times 10^{46}$ & $6.8 \times 10^{46}$ & -- & -- \\ \hline
%\multicolumn{7}{c}{}\\ \hline
% & & & & & & \\ \hline
C & IC-170922 & $8.6 \times 10^{-11}$ & -- & -- & -- & -- & $3.2 \times 10^{46}$ & $6.8 \times 10^{46}$ \\
\hline
\end{tabular}
%\end{minipage}
\end{center}
\tablenotetext{a}{A standard H$_0$=70, $\Omega_{M}$=0.3, $\Omega_{\Lambda}$=0.7 cosmology 
has been used here for the calculation of the corresponding luminosity distance.}
\end{table}

Table~\ref{table_luminosities} displays the gamma-ray energy density fluxes and luminosities 
above 100 MeV, estimated in a time interval of about $\pm 1$ week around T$_0$, 
and the neutrino luminosities estimated assuming a source active period 
of 6 months. Interestingly, for each of the three events 
we obtain similar values of the luminosities in gamma ray and neutrinos. 
The observed power for the two adopted distances (assumed to be emitted 
isotropically), is typical of Galactic and extragalactic compact objects 
being in the range of 10$^{36}$ erg s$^{-1}$ or 10$^{47}$ erg s$^{-1}$, respectively.

In case of IC-170922A (Source C) we observed a significant temporal correlation 
between the neutrino event and the almost simultaneous gamma-ray activity 
in HE and VHE bands from the IBL/HBL BL Lac type of blazar 
TXS 0506+056~\citep{2018Sci...361.1378I}. This is suggestive of this AGN 
sub-class of blazars being one of the main VHE neutrino emitters from hadronic processes. 
In the other two cases (A and B) there is no clear evidence of flaring activity 
from any known e.m. source inside the \agile/IceCube common error 
circles. Search for possible e.m. counterparts within the common \agile/IceCube 
error regions, initially focused on the identification of unknown HBL/HSP 
blazar candidates, found no obvious blazar candidates for Source A, as 
discussed in \cite{2017ApJ...846..121L}. A further investigation made 
in this work has identified a new possible HBL candidate, the 
elliptical galaxy SDSS J141457.72-002058.6. Regarding the gamma-ray Source B 
presented in this paper for the first time, some potential HBL blazars are found 
within the uncertainty neutrino/gamma regions. Moreover, a {\it FERMI}--3FGL 
source, 3FGL J0627.9-1517, recently associated to a HSP blazar, is at 
the boundary of the 90\% IceCube error box, although well outside 
the smaller AGILE error circle obtained on the longer integration around T$_0$ 
(see Fig.~\ref{fig3}). %(see Fig.~\ref{fig5}).

Given the lack of clear blazar counterparts for sources A and B, 
we are led to explore alternative explanations. Different classes 
of extragalactic sources, potentially hosting hadronic processes 
(bursts from radio-galaxies, starburst galaxies, eruptions from AGN cores, etc.) 
might be invoked to explain the gamma/neutrino correlations for A and B. 
Furthermore, IceCube neutrino fluxes can be produced also 
by “gamma-ray hidden sources” for which the high matter/radiation density 
surrounding a central engine enhances the target matter for the $p-p$ or $p-\gamma$ 
absorption processes. This would result in an observable neutrino flux with 
a highly suppressed gamma-ray flux from neutral pions’ decay. The \agile 
detections of gamma-ray sources near IC-160731 (Source A) and IC-170321 (Source B) 
indicate the possibility that, from time to time, under particularly favorable 
conditions the neutrino source may become transparent to MeV/GeV gamma-rays. 
Taking into account the optimized \agile sensitivity to soft gamma-ray emission 
in the 100--400 MeV energy band, sources with softer spectrum can also be considered. 
For example, in such a cases of enhanced target density, we might expect to 
observe a soft gamma-ray component peaking at MeV/sub-GeV due to the 
reprocessing of the VHE photons emitted by the pions’ decay.

The detection of the gamma-ray Source B within the IC-170321 error 
box is interesting. Its position is close to the Galactic plane with no clear 
extragalactic known gamma-ray counterpart. This source might be considered 
to belong to a class of neutrino sources possibly associated with a 
sub-dominant population of IceCube events apparently aligned near the 
Galactic plane~\citep{2017APh....86...46H}. Future observations 
will explore this very interesting possibility 
related to “hidden” compact objects in our Galaxy. We note that {\it FERMI}--LAT 
placed only a 95\% c.l. upper limit on the gamma-ray emission above 100 MeV 
over an interval of 1 week just before T$_0$~\citep{2017GCN.20971....1B}. 
Similar cases of \agile sources, both transients and steady, not 
confirmed by {\it FERMI}--LAT have been detected in the past~\citep{2009A&A...506.1563P, 
2013A&A...558A.137V, CAT2}. 
Several reasons can explain these discrepancies: source variability; 
different spectral response of the instruments; source visibility/exposure 
due to the different observing modes; \added{\edit1{event classification 
algorithms; background model (especially important for sources near the Galactic plane).}} 
All these factors may become important for relatively short gamma-ray 
transients \added{(with duration of a few days) at the level 
of 4$\sigma$ above the background}\footnote{The non-detection of 
the \agile Source A by {\it FERMI}--LAT was explained by a very poor 
visibility of the IC-160731 sky region in the days 
near T$_0$~\citep{2017ApJ...846..121L}.}.

This is the first time that evidence of multiple gamma-ray sources in 
close spatial and temporal coincidence with cosmic neutrinos is presented. 
\agile continues monitoring the gamma-ray sky and to react to IceCube alerts. 
More simultaneous neutrino and gamma-ray events are needed to strengthen 
the correlation indicated in the current \agile data analysis. From our analysis, 
different classes of neutrino sources should be considered. Continuous 
blazar monitoring is needed to confirm the association of BL Lac-type sources 
as in the case of our Source C and, in principle, galactic sources should be also 
taken into account and included in future searches.

Future studies of neutrino and gamma-ray sources need sensitive 
detectors and space missions able to reveal transient episodes 
occurring in the MeV/sub--GeV energy band. The e-ASTROGAM space 
mission~\citep{2017ExA....44...25D} has been proposed as well as 
the mission AMEGO, which is in an advanced state of development~\citep{2017HEAD...1610313M}. 
They can accomplish this task in the 2030's along with the upgraded 
neutrino experiment IceCube-Gen2~\citep{2015arXiv151005228T} and the 
new generations of gamma-ray and neutrino telescopes such as 
CTA and KM3NET~\citep{2013APh....43....3A, 2016JPhG...43h4001A}.

\acknowledgments
\agile is an ASI space mission developed with scientific and programmatic support 
from INAF and INFN. Research partially supported through the ASI grant no. I/028/12/2. 
We thank ASI personnel involved in the operations and data center of the \agile mission. 
Part of this work is based on archival data, software or online services provided by 
the Space Science Data Center (SSDC) -- ASI. It is also based on data and/or 
software provided by the High Energy Astrophysics Science Archive Research Center (HEASARC), 
which is a service of the Astrophysics Science Division at NASA/GSFC and the 
High Energy Astrophysics Division of the Smithsonian Astrophysical Observatory. 
This research has also made use of the SIMBAD database and the VizieR catalog 
access tool, operated at CDS, Strasbourg, France.
 
\software{{\agile} scientific analysis software (BUILD 21 \cite{AGILESW}), XIMAGE.}

\vspace{1cm}

\appendix
\section{Estimation of the False Alarm Rate (FAR) for {\it AGILE}--GRID 
in spinning mode}\label{FAR}
To evaluate the probability to find an \agile gamma-ray detection 
above 100 MeV in random coincidence with a candidate IceCube HESE/EHE 
astrophysical neutrino, we have estimated a {\it False Alarm Rate} (FAR) for 
the {\it AGILE}–-GRID data using the whole database of {\it Quick Look} (QL) 
detections hosted at the \agile Data Center.

Every day an automatic \agile QL procedure searches for gamma-ray transients 
above 100 MeV over the whole accessible sky~\citep{2014ApJ...781...19B}. The 
\agile QL runs since Nov. 2009, the starting of the spinning observation mode, 
over predefined data time intervals of 48 hours. Given the \agile effective area 
and sensitivity, these collecting time intervals are the most appropriate 
to accumulate enough statistics and to maximize the signal-to-noise ratio in spinning mode.

A blind search for gamma-ray transients is first applied to the data either 
using the {\it XIMAGE} detect algorithm or the so-called {\it spotfinder} 
method~\citep{2014ApJ...781...19B}. Then, counts, exposure and diffuse 
background model maps, centered at the excess positions found previously, 
are produced using the tasks of the \agile software, and eventually the count 
excess is evaluated against the expected background counts using the \agile 
Maximum Likelihood (ML) fit procedure~\citep{2012A&A...540A..79B}\footnote{For unknown 
sources, a simple power law spectral model with an index equal to -2.1 is 
usually assumed for the ML best-fit estimate procedure.}. All the gamma-ray 
detections and their ML best-estimate parameters (source significance as 
the square root of the ML Test Statistic ($\sqrt{TS}$), gamma-ray flux and 
source location) of each candidate source are then stored in the QL detection database.

\cite{2012A&A...540A..79B} assessed the \agile ML method, computing the 
chance probability to get a gamma-ray detection with a significance 
above a certain threshold, both for an empty \edit1{extragalactic} field and 
crowded Galactic fields. In our study, we need to extend that work in 
order to statistically determine the chance probability to have an \agile 
detection above a certain threshold of $\sqrt{TS}$ (over a 2-day time interval) 
{\it in temporal and spatial coincidence} with an IceCube neutrino event, having 
a localization error of the order of $1^\circ$ in radius.

Practically, to determine the FAR for the {\it AGILE}-–GRID data integrated 
over a 2-day interval, we proceed as following:

\begin{itemize}
\item a sky position in a relatively empty region of the \agile gamma-ray 
sky is considered and, using the QL database, the number of gamma-ray detections 
above a certain value of $\sqrt{TS}$ within a circular region of 20$^\circ$ 
in radius, centered at the chosen position, is counted;
\item the observed $\sqrt{TS}$ counting frequency is divided by the number 
of 1.5$^\circ$--radius pixels\footnote{Such pixel size is equal to the search 
cone radius used for the QL database scanning, which has been optimized 
according to the mean \agile angular resolution in the 100 MeV–-1 GeV 
energy band.} contained in the sky region under evaluation;
\item the $\sqrt{TS}$ counting frequency {\it per pixel} is divided by the \agile 
livetime computed since the beginning of the spinning mode (MJD=55139.5).
\end{itemize}

Since our minimum ``time unit'' is the 2-day integration time of 
the QL detections, the \agile livetime was expressed as the number 
of 2-day ``good'' maps generated since MJD=55139.5, i.e. having sufficient 
and uniform exposure to allow a reliable ML source parameter estimation. 

\added{\edit1{In this way, we basically end with a FAR for an \agile QL detection 
normalised to the solid angle subtended by a cone with half-aperture of 1.5$^\circ$ 
(i.e. the database search radius) and to the duration time of the QL maps. 
This FAR, expressed in units of 2-day maps and unit solid angle, can be then used 
to evaluate the probability of an accidental detection closed both in time 
and in space with an external event like the IceCube neutrinos.}}

\begin{figure}[t!]
\gridline{\leftfig{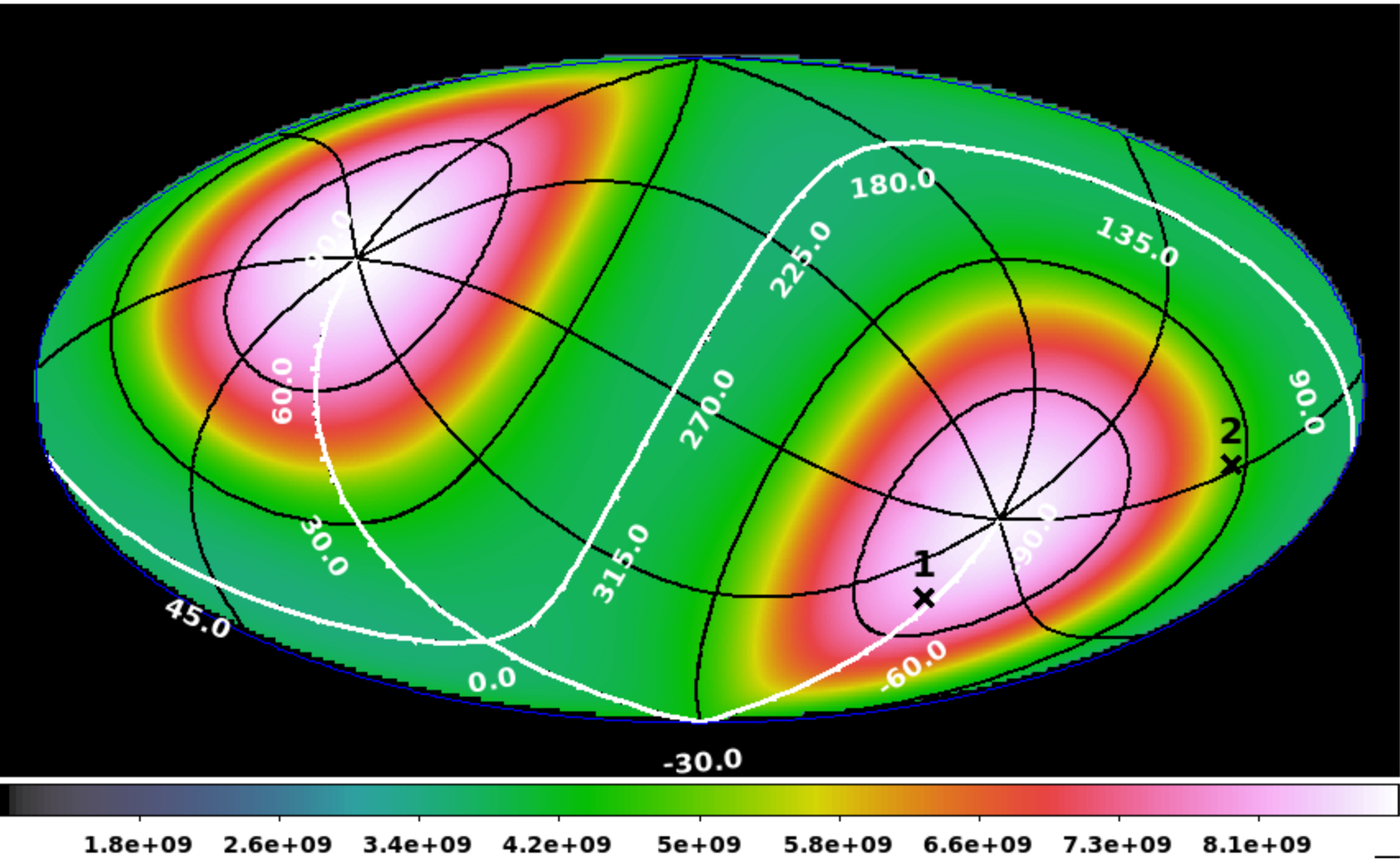}{0.5\textwidth}{}
          \rightfig{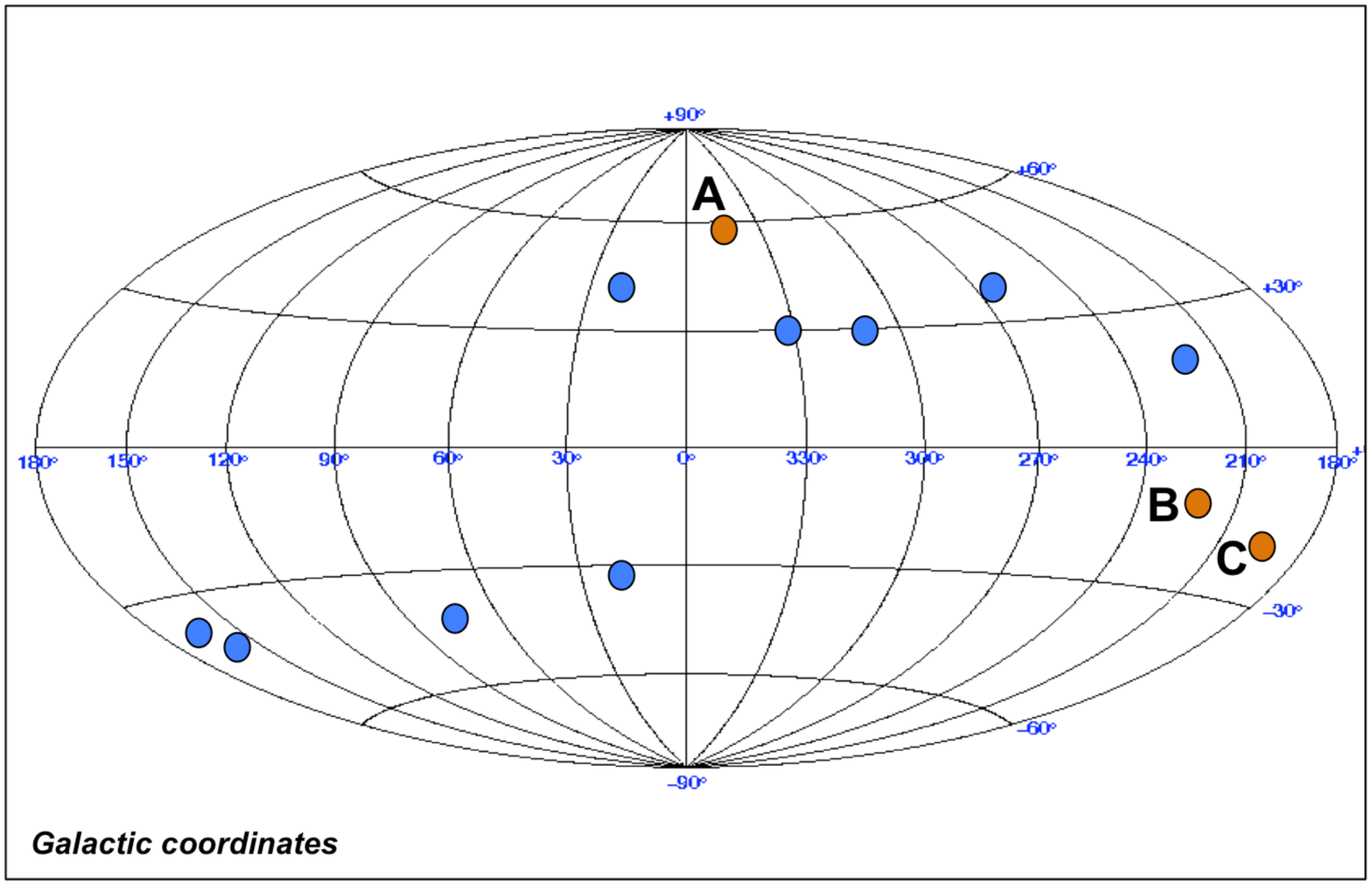}{0.53\textwidth}{}
          }
\vspace{-1.0cm}
\caption{{\it Left panel}: Hammer-Aitoff projection, in Galactic coordinates, of the total 
\agile gamma-ray exposure (in cm$^2$ s) accumulated since the beginning of the 
spinning observation mode (MJD=55139.5) up to the end of Sept. 2017 (MJD=58026.5). 
The overlaid grid defines the Ecliptic coordinate system. Due to the fixed orientation 
of the solar panels towards the Sun, the regions around the ecliptic poles are 
the most exposed, while the exposure along the ecliptic plane is affected 
by the apparent motion of the Sun/Anti-Sun exclusion regions. Positions labeled 
as {\it 1} and {\it 2} have been used to estimate the False Alarm Rate (FAR) 
for the {\it AGILE}-–GRID detections over 2-day time intervals.
{\it Right panel}: Distribution, in a Hammer-Aitoff projection in Galactic coordinates, 
of the reconstructed arrival directions of the IceCube HESE/EHE neutrino 
events published up to Aug. 2018. A, B, and C indicate the three events with 
an \agile \edit1{potential} gamma-ray counterpart.
\label{fig7}}
\end{figure}

The number of {\it 2-day good maps} varies according to the sky position 
considered, due both to the spacecraft rotation mode, in which the solar 
panels have to be kept fixed towards the Sun, and the seasonal variation 
of the Sun/Anti-Sun exclusion regions due to the Earth orbital 
motion\footnote{On average, the exclusion regions pass over the same sky 
position almost every three months.}. Figure~\ref{fig7}, {\it left panel}, shows 
an Hammer-Aitoff projection in Galactic coordinates of the overall \agile 
exposure (in cm$^2$ s), covering the period Nov. 2009 –- Sept. 2017 (MJD=55139.5 –- 58026.5). 
The regions around the Ecliptic poles are the most exposed, while the exposure along the Ecliptic 
plane is affected by the apparent motion of the Sun/Anti-Sun exclusion regions.

The total \agile livetime for the whole spinning period of almost 8 years, 
expressed in terms of total number of {\it 2-day good maps}, ranges from 
around 1000 for the regions near the Ecliptic poles down to around 200 for the 
less exposed sky positions on the Ecliptic plane\footnote{The number of 
{\it 2-day good maps} accumulated over the whole spinning period for the 
two positions considered has been estimated using the \agile online interactive 
analysis tool based on the {\it AGILE}-–GRID Level-3 (LV3) archive of 
pre-computed counts, exposure, and diffuse background emission maps 
(available at the URL: http://www.asdc.asi.it/mmia/index.php?mission=agilelv3mmia).}. 
For the FAR calculation, we considered a relatively empty region of brilliant 
gamma-ray sources placed near the south Ecliptic pole (position {\it 1} in 
Fig.~\ref{fig7}, {\it left panel}). For this position, the number 
of {\it 2-day good maps} amounts to 1000. The value of FAR estimated on 
position {\it 1} can be applied to the estimation 
of the false alarm probability in case of an \agile QL detection consistent 
with an IceCube HESE/EHE event laying well above the Galactic plane 
(with Galactic latitude $|b| \geq 20^\circ$).
 
Since \agile Sources B and C described in the main text are consistent 
with two IceCube events located nearer the Galactic plane ($b$=-10.75 and -19.56 (deg), 
respectively), to estimate the chance correlation for this region 
we considered a $20^\circ$ region centered at Galactic coordinates $l,b$=(217.0, -15) (deg) 
(position {\it 2} in Fig.~\ref{fig7}, {\it left panel}). Due to higher diffuse gamma-ray emission 
near the plane, the FAR for this region resulted to be roughly a 30\% higher 
than the value obtained at higher or lower Galactic latitudes, ending in a 
slightly higher value of the {\it post-trial} false alarm probability $P_i$ 
for the two events, as shown in Table~\ref{table_agile_detections}.

We notice that the FAR (and the false alarm probability) can be over-estimated by 20-30\% 
due to the presence of an un-subtracted non-Poissonian component of {\it real} 
gamma-ray transients from unknown sources occurring in the extraction sky region.

\vspace{0.5cm}

\section{IceCube HESE/EHE events announced since April 2016}\label{hese_list}
Table~\ref{ICECUBE_ALERTS} shows all the IceCube HESE/EHE events published up 
to Aug. 2018. Since Apr. 2016, these events are announced through the GCN/AMON notice 
circuit~\citep{2017APh....92...30A}, usually followed by a GCN Circular 
reporting the results of a further refined data analysis which provides 
improved reconstructed neutrino arrival directions and position uncertainties. 

Along with the IceCube event ID, the table shows:

\begin{itemize}
\setlength{\itemsep}{0pt}
\item the neutrino event time (in UT and MJD date); 
\item the event classification (HESE or EHE); 
\item the best fitted reconstructed neutrino arrival direction in Equatorial coordinates  
(J2000) and its uncertainty;
\item the corresponding Galactic coordinates l and b;
\item the GCN Circular number reporting about the refined analysis (if available).
\end{itemize}

Where available, the table shows the refined arrival direction published 
in the GCN Circular.

Event numbered 34032434 has been rejected after refined analysis~\citep{2017GCN.22065....1I} 
while events 65274589 and 32674593 were considered consistent with rare atmospheric 
muon background events~\citep{2017GCN.20857....1B, GCN21075} and, thus, 
they were not considered in our analysis.

\added{\edit1{Figure~\ref{fig7}, {\it right panel}, shows the distribution of 
all IceCube events in a Hammer-Aitoff projection of the sky in Galactic coordinates. 
All events appear well above the Galactic plane, except for one case 
(IC-170321) which shows a Galactic latitude of -10.75 degrees. 
The three neutrino events with an \agile possible transient 
counterpart, A, B, and C, are shown in orange.}}

%%Table Icecube alerts
\begin{longrotatetable}
\begin{deluxetable*}{lccccccccc}
\tablecaption{Public IceCube HESE/EHE alerts published through the GCN/AMON and GCN Circular network, and their main identification parameters.\label{ICECUBE_ALERTS}}
\tablewidth{500pt}
\tabletypesize{\tiny}
\tablehead{
\colhead{Event ID} & \colhead{Date Time} & \colhead{MJD} & \colhead{Event class} & 
\colhead{R.A. (J2000)} & \colhead{Decl. (J2000)} & \colhead{Position uncertainty} & 
\colhead{$l$} & \colhead{$b$} & \colhead{GCN Circ.} \\ 
\colhead{} & \colhead{(UT)} & \colhead{(days)} & \colhead{} & 
\colhead{(deg)} & \colhead{(deg)} & \colhead{(arcmin)} &
\colhead{(deg)} & \colhead{(deg)} & \colhead{\#}
}
\startdata
67093193 (IC-160427) &  16/04/27 05:52:32.00 & 57505.245 & HESE     & 240.57 & 9.34 & 36 (90\% c.r.) & 20.69 & 41.68 & 19363 \\
\hline
6888376 (IC-160731)  &  16/07/31 01:55:04.00 & 57600.079 & HESE/EHE & 214.544 & -0.3347 & 45 (90\% c.r.) & 343.68 & 55.52 & -- \\
\hline
26552458 (IC-160806) &  16/08/06 12:21:33.00 & 57606.515 & EHE      & 122.81  & -0.8061 & 34 (50\% c.r.) & 223.07 & 17.29 & 19787 \\
\hline
58537957 (IC-160814) &  16/08/14 21:45:54.00 & 57614.907 & HESE     & 199.31  & -32.0165& 89.4 (90\% c.r.) & 309.28 & 30.54 & -- \\
\hline
38561326 (IC-161103) &  16/11/03 09:07:31.12 & 57695.38  & HESE     & 40.8252  & 12.5592& 66 (90\% c.r.) & 160.90 & -41.92 & -- \\
\hline
80127519 (IC-161210) &  16/12/10 20:06:40.31 & 57732.838 & EHE      & 46.5799  & 14.98  & 60 (50\% c.r.) & 164.89 & -36.67 & -- \\
\hline
65274589 (IC-170312)\footnote{Possibly consistent with rare atmospheric muon background event.\label{atmnu}} & 17/03/12 13:49:39.83 & 57824.576 & HESE     & 305.15 & -26.61  & \shortstack{ \\ $\pm$30 in R.A. \\ $\pm$30 in Decl. \\ (90\% PSF)} & 16.50 & -30.40 & 20857 \\
\hline
80305071 (IC-170321) &  17/03/21 07:32:20.69 & 57833.314 & EHE      & 98.3     & -15.02  &  \shortstack{ \\ $\pm$72 in R.A. \\ $\pm$72 in Decl. \\ (90\% PSF)} & 224.42 & -10.75 & 20929 \\
\hline
32674593 (IC-170506)\textsuperscript{\ref{atmnu}} & 17/05/06 12:36:55.80 & 57879.526 & HESE    & 221.8    &  -26    & \shortstack{ \\ $\pm$180 in R.A. \\ $\pm$120 in Decl. \\ (90\% PSF)} & 332.95 & 30.03 & 21075 \\
\hline
50579430 (IC-170922) &  17/09/22 20:54:30.43 & 58018.871 & EHE     &  77.43   &   5.72  & \shortstack{ \\ -48/+78 in R.A. \\ -24/+42 in Decl. \\ (90\% PSF)} & 195.42 & -19.56 & 21916 \\
\hline
56068624 (IC-171015) & 17/10/15 01:34:30.06 & 58041.066 & HESE    & 162.86   &  -15.44 & \shortstack{ \\ -102/+156 in R.A. \\ -120/+96 in Decl. \\ (90\% PSF)} & 264.96 &  38.43 & 22016 \\
\hline
34032434 (IC-171028)\footnote{Event candidate retracted in GCN Circular \#22065.} &  17/10/28 08:28:14.81 &   58054.353 & HESE    & 275.076  &  34.5011&  -- &  61.96 &  20.95 & 22065 \\ 
\hline
17569642 (IC-171106)\footnote{Event is consistent with being produced by a neutrino with energy in excess of 1~PeV.} &  17/11/06 18:39:39.21 &   58063.778 & EHE     & 340.0 & 7.4 & \shortstack{ \\ -30/+42 in R.A. \\ -15/+21 in Decl. \\ (90\% PSF)} & 75.51 & -43.05 & 22105 \\
\enddata
\end{deluxetable*}
\end{longrotatetable}

\clearpage

\bibliographystyle{aasjournal}
\bibliography{Lucarelli_AGILE-ICECUBEevents_astroph}

\begin{thebibliography}{}
\expandafter\ifx\csname natexlab\endcsname\relax\def\natexlab#1{#1}\fi
\providecommand{\url}[1]{\href{#1}{#1}}
\providecommand{\dodoi}[1]{doi:~\href{http://doi.org/#1}{\nolinkurl{#1}}}
\providecommand{\doeprint}[1]{\href{http://ascl.net/#1}{\nolinkurl{http://ascl.net/#1}}}
\providecommand{\doarXiv}[1]{\href{https://arxiv.org/abs/#1}{\nolinkurl{https://arxiv.org/abs/#1}}}

\bibitem[{{Aartsen} {et~al.}(2017{\natexlab{a}}){Aartsen}, {Abraham},
  {Ackermann}, {Adams}, {et~al.}}]{2017ApJ...835..151A}
{Aartsen}, M.~G., {Abraham}, K., {Ackermann}, M., {Adams}, J., {et~al.}
  2017{\natexlab{a}}, \apj, 835, 151, \dodoi{10.3847/1538-4357/835/2/151}

\bibitem[{{Aartsen} {et~al.}(2013){Aartsen}, {Abraham}, {Ackermann},
  {et~al.}}]{2013Sci...342E...1I}
{Aartsen}, M.~G., {Abraham}, K., {Ackermann}, M., {et~al.} 2013, Science, 342,
  1242856, \dodoi{10.1126/science.1242856}

\bibitem[{{Aartsen} {et~al.}(2017{\natexlab{b}}){Aartsen}, {Abraham},
  {Ackermann}, {et~al.}}]{2017ApJ...835...45A}
---. 2017{\natexlab{b}}, \apj, 835, 45, \dodoi{10.3847/1538-4357/835/1/45}

\bibitem[{{Aartsen} {et~al.}(2017{\natexlab{c}}){Aartsen}, {Ackermann},
  {Adams}, {et~al.}}]{2017APh....92...30A}
{Aartsen}, M.~G., {Ackermann}, M., {Adams}, J., {et~al.} 2017{\natexlab{c}},
  APh, 92, 30, \dodoi{10.1016/j.astropartphys.2017.05.002}

\bibitem[{{Aartsen} {et~al.}(2018{\natexlab{a}}){Aartsen}, {Ackermann},
  {Adams}, {et~al.}}]{2018Sci...361.1378I}
---. 2018{\natexlab{a}}, Science, 361, eaat1378,
  \dodoi{10.1126/science.aat1378}

\bibitem[{{Aartsen} {et~al.}(2018{\natexlab{b}}){Aartsen}, {Ackermann},
  {Adams}, {et~al.}}]{2018Sci...361..147I}
---. 2018{\natexlab{b}}, Science, 361, 147, \dodoi{10.1126/science.aat2890}

\bibitem[{{Aartsen} {et~al.}(2015){Aartsen}, {Abraham}, {Ackermann}, {Adams},
  {Aguilar}, {Ahlers}, {Ahrens}, {Altmann}, {Anderson}, {Archinger},
  {et~al.}}]{2015PhRvL.115h1102A}
{Aartsen}, M.~G., {Abraham}, K., {Ackermann}, M., {et~al.} 2015, Physical
  Review Letters, 115, 081102, \dodoi{10.1103/PhysRevLett.115.081102}

\bibitem[{{Acero} {et~al.}(2015){Acero}, {Ackermann}, {Ajello},
  {et~al.}}]{2015ApJS..218...23A}
{Acero}, F., {Ackermann}, M., {Ajello}, M., {et~al.} 2015, Ap. J. Supp., 218,
  23, \dodoi{10.1088/0067-0049/218/2/23}

\bibitem[{{Acharya} {et~al.}(2013){Acharya}, {Actis}, {Aghajani}, {Agnetta},
  {et~al.}}]{2013APh....43....3A}
{Acharya}, B.~S., {Actis}, M., {Aghajani}, T., {Agnetta}, G., {et~al.} 2013,
  Astroparticle Physics, 43, 3, \dodoi{10.1016/j.astropartphys.2013.01.007}

\bibitem[{{Adri{\'a}n-Mart{\'{\i}}nez}
  {et~al.}(2016){Adri{\'a}n-Mart{\'{\i}}nez}, {Ageron}, {Aharonian},
  {et~al.}}]{2016JPhG...43h4001A}
{Adri{\'a}n-Mart{\'{\i}}nez}, S., {Ageron}, M., {Aharonian}, F., {et~al.} 2016,
  Journal of Physics G Nuclear Physics, 43, 084001,
  \dodoi{10.1088/0954-3899/43/8/084001}

\bibitem[{{Ahlers} {et~al.}(2016){Ahlers}, {Bai}, {Barger}, \&
  {Lu}}]{2016PhRvD..93a3009A}
{Ahlers}, M., {Bai}, Y., {Barger}, V., \& {Lu}, R. 2016, \prd, 93, 013009,
  \dodoi{10.1103/PhysRevD.93.013009}

\bibitem[{{Albert} {et~al.}(2017){Albert}, {Andr{\'e}}, {Anghinolfi}, {Anton},
  {et~al.}}]{2017PhRvD..96h2001A}
{Albert}, A., {Andr{\'e}}, M., {Anghinolfi}, M., {Anton}, G., {et~al.} 2017,
  \prd, 96, 082001, \dodoi{10.1103/PhysRevD.96.082001}

\bibitem[{{Anchordoqui} {et~al.}(2014){Anchordoqui}, {Barger}, {Cholis},
  {Goldberg}, {Hooper}, {Kusenko}, {Learned}, {Marfatia}, {Pakvasa}, {Paul}, \&
  {Weiler}}]{Anchordoqui:2013dnh}
{Anchordoqui}, L.~A., {Barger}, V., {Cholis}, I., {et~al.} 2014, Journal of
  High Energy Astrophysics, 1, 1, \dodoi{10.1016/j.jheap.2014.01.001}

\bibitem[{{Becker Tjus} {et~al.}(2014){Becker Tjus}, {Eichmann}, {Halzen},
  {Kheirandish}, \& {Saba}}]{2014PhRvD..89l3005B}
{Becker Tjus}, J., {Eichmann}, B., {Halzen}, F., {Kheirandish}, A., \& {Saba},
  S.~M. 2014, \prd, 89, 123005, \dodoi{10.1103/PhysRevD.89.123005}

\bibitem[{{Bednarek}(2005)}]{2005ApJ...631..466B}
{Bednarek}, W. 2005, ApJ, 631, 466, \dodoi{10.1086/432411}

\bibitem[{{Blaufuss}(2017{\natexlab{a}})}]{2017GCN.20929....1B}
{Blaufuss}, E. 2017{\natexlab{a}}, GRB Coordinates Network, Circular Service,
  No.~20929, \#1 (2017), 20929

\bibitem[{{Blaufuss}(2017{\natexlab{b}})}]{GCN21916}
---. 2017{\natexlab{b}}, {IceCube-170922A - IceCube observation of a
  high-energy neutrino candidate event.}

\bibitem[{{Blaufuss}(2017{\natexlab{c}})}]{2017GCN.20857....1B}
---. 2017{\natexlab{c}}, GRB Coordinates Network, Circular Service, No.~20857,
  \#1 (2017), 20857

\bibitem[{{Blaufuss}(2017{\natexlab{d}})}]{GCN21075}
---. 2017{\natexlab{d}}, {IceCube-170506A - IceCube update on a high-energy
  neutrino candidate event}

\bibitem[{{Bulgarelli} {et~al.}(2012){Bulgarelli}, {Chen}, {Tavani},
  {Gianotti}, {Trifoglio}, \& {Contessi}}]{2012A&A...540A..79B}
{Bulgarelli}, A., {Chen}, A.~W., {Tavani}, M., {et~al.} 2012, A\&A, 540, A79,
  \dodoi{10.1051/0004-6361/201118023}

\bibitem[{{Bulgarelli} {et~al.}(2018){Bulgarelli}, {Fioretti}, {Parmiggiani},
  {et~al.}}]{CAT2}
{Bulgarelli}, A., {Fioretti}, V., {Parmiggiani}, N., {et~al.} 2018, {Submitted
  for publication in ApJ.}

\bibitem[{{Bulgarelli} {et~al.}(2014){Bulgarelli}, {Trifoglio}, {Gianotti},
  {et~al.}}]{2014ApJ...781...19B}
{Bulgarelli}, A., {Trifoglio}, M., {Gianotti}, F., {et~al.} 2014, ApJ, 781, 19,
  \dodoi{10.1088/0004-637X/781/1/19}

\bibitem[{{Buson} {et~al.}(2017){Buson}, {Kocevski}, \&
  {Ciprini}}]{2017GCN.20971....1B}
{Buson}, S., {Kocevski}, D., \& {Ciprini}, S. 2017, GRB Coordinates Network,
  Circular Service, No.~20971, \#1 (2017), 20971

\bibitem[{{Chang} {et~al.}(2017){Chang}, {Arsioli}, {Giommi}, \&
  {Padovani}}]{2017A&A...598A..17C}
{Chang}, Y.-L., {Arsioli}, B., {Giommi}, P., \& {Padovani}, P. 2017, A\&A, 598,
  A17, \dodoi{10.1051/0004-6361/201629487}

\bibitem[{{Chang} {et~al.}(2018){Chang}, {Arsioli}, {Giommi}, {Padovani}, \&
  {Brandt}}]{3HSP}
{Chang}, Y.-L., {Arsioli}, B., {Giommi}, P., {Padovani}, P., \& {Brandt}, C.
  2018, {Submitted for publication in A\&A.}

\bibitem[{{Chen} {et~al.}(2011){Chen}, {Bulgarelli}, {Contessi}, {GIULIANI},
  {MEREGHETTI}, {PELLIZZONI}, {TROIS}, \& {VERCELLONE}}]{AGILESW}
{Chen}, A.~W., {Bulgarelli}, A., {Contessi}, T., {et~al.} 2011, {GRID}
  Scientific Analysis -- {USER MANUAL},  {\tt http://agile.asdc.asi.it}

\bibitem[{{Connaughton} {et~al.}(2016){Connaughton}, {Burns}, {Goldstein},
  {Blackburn}, {Briggs}, {Zhang}, {Camp}, {Christensen}, {Hui}, {Jenke},
  {Littenberg}, {McEnery}, {Racusin}, {Shawhan}, {Singer}, {Veitch},
  {Wilson-Hodge}, {Bhat}, {Bissaldi}, {Cleveland}, {Fitzpatrick}, {Giles},
  {Gibby}, {von Kienlin}, {Kippen}, {McBreen}, {Mailyan}, {Meegan}, {Paciesas},
  {Preece}, {Roberts}, {Sparke}, {Stanbro}, {Toelge}, \&
  {Veres}}]{2016ApJ...826L...6C}
{Connaughton}, V., {Burns}, E., {Goldstein}, A., {et~al.} 2016, ApJL, 826, L6,
  \dodoi{10.3847/2041-8205/826/1/L6}

\bibitem[{{De Angelis} {et~al.}(2017){De Angelis}, {Tatischeff}, {Tavani},
  {Oberlack}, {et~al.}}]{2017ExA....44...25D}
{De Angelis}, A., {Tatischeff}, V., {Tavani}, M., {Oberlack}, U., {et~al.}
  2017, Experimental Astronomy, 44, 25, \dodoi{10.1007/s10686-017-9533-6}

\bibitem[{{Gaisser} {et~al.}(1995){Gaisser}, {Halzen}, \&
  {Stanev}}]{1995PhR...258..173G}
{Gaisser}, T.~K., {Halzen}, F., \& {Stanev}, T. 1995, \physrep, 258, 173,
  \dodoi{10.1016/0370-1573(95)00003-Y}

\bibitem[{{Halzen}(2017)}]{2017NatPh..13..232H}
{Halzen}, F. 2017, Nature Physics, 13, 232, \dodoi{10.1038/nphys3816}

\bibitem[{{Halzen} {et~al.}(2017){Halzen}, {Kheirandish}, \&
  {Niro}}]{2017APh....86...46H}
{Halzen}, F., {Kheirandish}, A., \& {Niro}, V. 2017, Astroparticle Physics, 86,
  46, \dodoi{10.1016/j.astropartphys.2016.11.004}

\bibitem[{{Icecube Collaboration}(2017)}]{2017GCN.22065....1I}
{Icecube Collaboration}. 2017, GRB Coordinates Network, 22065, 1

\bibitem[{{Keivani}(2017)}]{GCN20964}
{Keivani}, A. 2017, {IceCube-170321A: Swift-XRT observations.}

\bibitem[{{Kotera} {et~al.}(2009){Kotera}, {Allard}, {Murase}, {Aoi}, {Dubois},
  {Pierog}, \& {Nagataki}}]{2009ApJ...707..370K}
{Kotera}, K., {Allard}, D., {Murase}, K., {et~al.} 2009, \apj, 707, 370,
  \dodoi{10.1088/0004-637X/707/1/370}

\bibitem[{{Lamastra} {et~al.}(2017){Lamastra}, {Menci}, {Fiore}, {Antonelli},
  {Colafrancesco}, {Guetta}, \& {Stamerra}}]{2017A&A...607A..18L}
{Lamastra}, A., {Menci}, N., {Fiore}, F., {et~al.} 2017, \aap, 607, A18,
  \dodoi{10.1051/0004-6361/201731452}

\bibitem[{{Loeb} \& {Waxman}(2006)}]{2006JCAP...05..003L}
{Loeb}, A., \& {Waxman}, E. 2006, \jcap, 5, 003,
  \dodoi{10.1088/1475-7516/2006/05/003}

\bibitem[{{Lucarelli} {et~al.}(2017{\natexlab{a}}){Lucarelli}, {Piano},
  {Pittori}, {Verrecchia}, {et~al.}}]{2017ATel10801....1L}
{Lucarelli}, F., {Piano}, G., {Pittori}, C., {Verrecchia}, F., {et~al.}
  2017{\natexlab{a}}, The Astronomer's Telegram, 10801

\bibitem[{{Lucarelli} {et~al.}(2014){Lucarelli}, {Pittori}, {Verrecchia},
  {Tavani}, {et~al.}}]{2014ATel.6457....1L}
{Lucarelli}, F., {Pittori}, C., {Verrecchia}, F., {Tavani}, M., {et~al.} 2014,
  The Astronomer's Telegram, 6457

\bibitem[{{Lucarelli} {et~al.}(2017{\natexlab{b}}){Lucarelli}, {Pittori},
  {Verrecchia}, {Donnarumma}, {Tavani}, {Bulgarelli}, {Giuliani}, {Antonelli},
  {Caraveo}, {Cattaneo}, {Colafrancesco}, {Longo}, {Mereghetti}, {Morselli},
  {Pacciani}, {Piano}, {Pellizzoni}, {Pilia}, {Rappoldi}, {Trois}, \&
  {Vercellone}}]{2017ApJ...846..121L}
{Lucarelli}, F., {Pittori}, C., {Verrecchia}, F., {et~al.} 2017{\natexlab{b}},
  ApJ, 846, 121, \dodoi{10.3847/1538-4357/aa81c8}

\bibitem[{{Mannheim}(1995)}]{1995APh.....3..295M}
{Mannheim}, K. 1995, APh, 3, 295, \dodoi{10.1016/0927-6505(94)00044-4}

\bibitem[{{Massaro} {et~al.}(2015){Massaro}, {Maselli}, {Leto}, {Marchegiani},
  {Perri}, {Giommi}, \& {Piranomonte}}]{2015Ap&SS.357...75M}
{Massaro}, E., {Maselli}, A., {Leto}, C., {et~al.} 2015, Astrophysics and Space
  Science, 357, 75, \dodoi{10.1007/s10509-015-2254-2}

\bibitem[{{McEnery}(2017)}]{2017HEAD...1610313M}
{McEnery}, J.~E. 2017, in AAS/High Energy Astrophysics Division, Vol.~16,
  AAS/High Energy Astrophysics Division \#16, 103.13

\bibitem[{{M{\'e}sz{\'a}ros}(2017)}]{2017ARNPS..67...45M}
{M{\'e}sz{\'a}ros}, P. 2017, Annual Review of Nuclear and Particle Science, 67,
  45, \dodoi{10.1146/annurev-nucl-101916-123304}

\bibitem[{{Murase} {et~al.}(2008){Murase}, {Inoue}, \&
  {Nagataki}}]{2008ApJ...689L.105M}
{Murase}, K., {Inoue}, S., \& {Nagataki}, S. 2008, \apjl, 689, L105,
  \dodoi{10.1086/595882}

\bibitem[{{Ojha} {et~al.}(2014){Ojha}, {Carpenter}, {Becerra}, {Krauss},
  {et~al.}}]{2014ATel.6425....1O}
{Ojha}, R., {Carpenter}, B., {Becerra}, J., {Krauss}, F., {et~al.} 2014, The
  Astronomer's Telegram, 6425

\bibitem[{{Padovani} {et~al.}(2018){Padovani}, {Giommi}, {Resconi}, {Glauch},
  {Arsioli}, {Sahakyan}, \& {Huber}}]{2018MNRAS.480..192P}
{Padovani}, P., {Giommi}, P., {Resconi}, E., {et~al.} 2018, \mnras, 480, 192,
  \dodoi{10.1093/mnras/sty1852}

\bibitem[{{Padovani} {et~al.}(2016){Padovani}, {Resconi}, {Giommi}, {Arsioli},
  \& {Chang}}]{2016MNRAS.457.3582P}
{Padovani}, P., {Resconi}, E., {Giommi}, P., {Arsioli}, B., \& {Chang}, Y.~L.
  2016, Mon. Not. R. Astron. Soc., 457, 3582, \dodoi{10.1093/mnras/stw228}

\bibitem[{{Paiano} {et~al.}(2018){Paiano}, {Falomo}, {Treves}, \&
  {Scarpa}}]{2018ApJ...854L..32P}
{Paiano}, S., {Falomo}, R., {Treves}, A., \& {Scarpa}, R. 2018, \apjl, 854,
  L32, \dodoi{10.3847/2041-8213/aaad5e}

\bibitem[{{Pittori} {et~al.}(2009){Pittori}, {Verrecchia}, {Chen},
  {et~al.}}]{2009A&A...506.1563P}
{Pittori}, C., {Verrecchia}, F., {Chen}, A.~W., {et~al.} 2009, \aap, 506, 1563,
  \dodoi{10.1051/0004-6361/200911783}

\bibitem[{{Resconi} {et~al.}(2017){Resconi}, {Coenders}, {Padovani}, {Giommi},
  \& {Caccianiga}}]{2017MNRAS.468..597R}
{Resconi}, E., {Coenders}, S., {Padovani}, P., {Giommi}, P., \& {Caccianiga},
  L. 2017, Mon. Not. R. Astron. Soc., 468, 597, \dodoi{10.1093/mnras/stx498}

\bibitem[{{Sabatini} {et~al.}(2015){Sabatini}, {Donnarumma}, {Tavani},
  {et~al.}}]{2015ApJ...809...60S}
{Sabatini}, S., {Donnarumma}, I., {Tavani}, M., {et~al.} 2015, ApJ, 809, 60,
  \dodoi{10.1088/0004-637X/809/1/60}

\bibitem[{{Sahakyan} {et~al.}(2014){Sahakyan}, {Piano}, \&
  {Tavani}}]{2014ApJ...780...29S}
{Sahakyan}, N., {Piano}, G., \& {Tavani}, M. 2014, ApJ, 780, 29,
  \dodoi{10.1088/0004-637X/780/1/29}

\bibitem[{{Savchenko} {et~al.}(2017){Savchenko}, {Santander}, {Keivani},
  {et~al.}}]{2017GCN.20937....1S}
{Savchenko}, V., {Santander}, M., {Keivani}, A., {et~al.} 2017, GRB Coordinates
  Network, Circular Service, No.~20937, \#1 (2017), 20937

\bibitem[{{Svinkin} {et~al.}(2017){Svinkin}, {Golenetskii}, {Aptekar},
  {et~al.}}]{2017GCN.20973....1S}
{Svinkin}, D., {Golenetskii}, S., {Aptekar}, R., {et~al.} 2017, GRB Coordinates
  Network, Circular Service, No.~20973, \#1 (2017), 20973

\bibitem[{{Tanaka} {et~al.}(2017){Tanaka}, {Buson}, \&
  {Kocevski}}]{2017ATel10791....1T}
{Tanaka}, Y.~T., {Buson}, S., \& {Kocevski}, D. 2017, The Astronomer's
  Telegram, 10791

\bibitem[{{Tavani} {et~al.}(2009){Tavani}, {Barbiellini}, {Argan},
  {et~al.}}]{2009A&A...502..995T}
{Tavani}, M., {Barbiellini}, G., {Argan}, A., {et~al.} 2009, A\&A, 502, 995,
  \dodoi{10.1051/0004-6361/200810527}

\bibitem[{{Tavecchio} {et~al.}(2018){Tavecchio}, {Righi}, {Capetti}, {Grandi},
  \& {Ghisellini}}]{2018MNRAS.tmp..247T}
{Tavecchio}, F., {Righi}, C., {Capetti}, A., {Grandi}, P., \& {Ghisellini}, G.
  2018, \mnras, \dodoi{10.1093/mnras/sty251}

\bibitem[{{The IceCube-Gen2 Collaboration} {et~al.}(2015){The IceCube-Gen2
  Collaboration}, {:}, {Aartsen}, {Abraham}, {Ackermann}, {Adams},
  {et~al.}}]{2015arXiv151005228T}
{The IceCube-Gen2 Collaboration}, {:}, {Aartsen}, M.~G., {et~al.} 2015, ArXiv
  e-prints.
\newblock \doarXiv{1510.05228}

\bibitem[{{Verrecchia} {et~al.}(2013){Verrecchia}, {Pittori}, {Chen},
  {et~al.}}]{2013A&A...558A.137V}
{Verrecchia}, F., {Pittori}, C., {Chen}, A.~W., {et~al.} 2013, \aap, 558, A137,
  \dodoi{10.1051/0004-6361/201321452}

\bibitem[{{Vissani}(2006)}]{2006APh....26..310V}
{Vissani}, F. 2006, Astroparticle Physics, 26, 310,
  \dodoi{10.1016/j.astropartphys.2006.07.005}

\bibitem[{{Wang} \& {Loeb}(2016)}]{2016JCAP...12..012W}
{Wang}, X., \& {Loeb}, A. 2016, \jcap, 12, 012,
  \dodoi{10.1088/1475-7516/2016/12/012}

\end{thebibliography}



\end{document}